\documentclass[
12pt]{article}
\usepackage{color}
\usepackage{graphicx}
 \usepackage{amsmath,amssymb,array,calc}
\usepackage{bm}

\definecolor{darkgreen}{rgb}{0,0.55,0}

\newcommand{\bea}{\begin{eqnarray}}
\newcommand{\eea}{\end{eqnarray}}
\newcommand{\be}{\begin{equation}}
\newcommand{\ee}{\end{equation}}
\newcommand{\nn}{\nonumber}

\def\p{\phi}

\def\revise#1       {\raisebox{-0em}{\rule{3pt}{1em}}%
                     \marginpar{\raisebox{.5em}{\vrule width3pt\
                     \vrule width0pt height 0pt depth0.5em
                     \hbox to 0cm{\hspace{0cm}{%
                     \parbox[t]{4em}{\raggedright\footnotesize{#1}}}\hss}}}}

\def\gomega       {{\bm \omega}}
\def\p            {{\bm p}}
\def\gOmega       {{\bm \Omega}}
\def\M            {{\bm M}}
\def\hd           {{\hat d}}

\def\sqr#1#2{{\vcenter{\vbox{\hrule height.#2pt
 \hbox{\vrule width.#2pt height#1pt \kern#1pt
 \vrule width.#2pt}\hrule height.#2pt}}}}

\long\def\symbolfootnote[#1]#2{\begingroup%
\def\thefootnote{\fnsymbol{footnote}}\footnote[#1]{#2}\endgroup} 

\catcode`\@=12

\begin{document}

\makeatletter \@addtoreset{equation}{section} \makeatother
\renewcommand{\theequation}{\thesection.\arabic{equation}}

\renewcommand\baselinestretch{1.25}
\setlength{\paperheight}{10in}
\setlength{\paperwidth}{9.5in}

\begin{titlepage}


%

\vskip 1cm



\vskip 3cm

\begin{center}
{\bf \large Holographic real-time non-relativistic correlators at zero and finite temperature}
\end{center}
\vskip 1cm

\centerline{{\large  Edwin Barnes\symbolfootnote[1]{Present address: CMTC, University of Maryland, College Park, MD 20742, barnes@umd.edu} Diana Vaman\symbolfootnote[2]{dv3h@virginia.edu}, Chaolun Wu\symbolfootnote[3]{cw2an@virginia.edu}}
}

\vskip .5cm
\centerline{\it Department of Physics, The University of Virginia}
\centerline{\it McCormick Rd, Charlottesville, VA 22904}

\vspace{1cm}

\begin{abstract}

We compute a variety of two and three-point real-time correlation functions for a strongly-coupled non-relativistic field theory. We focus on the theory conjectured to be dual to the Schr\"{o}dinger-invariant gravitational spacetime introduced by Balasubramanian, McGreevy, and Son, but our methods apply to a large class of non-relativistic theories. At zero temperature, we obtain time-ordered, retarded, and Wightman non-relativistic correlators for scalar operators of arbitrary conformal dimension directly in field theory by applying a certain lightlike Fourier transform to relativistic conformal correlators, and we verify that non-relativistic AdS/CFT reproduces the results. We compute thermal two and three-point real-time correlators for scalar operators dual to scalar fields in the black hole background which is the finite temperature generalization of the Schr\"{o}dinger spacetime. This is done by first identifying thermal real-time bulk-to-boundary propagators which, when combined with Veltman's circling rules, yield two and three-point correlators. The two-point correlators we obtain satisfy the Kallen-Lehmann relations. We also give retarded and time-ordered three-point correlators.

\end{abstract}

\end{titlepage}



\tableofcontents\eject

\section{Introduction and Summary}
The AdS/CFT correspondence \cite{Maldacena:1997re,Witten:1998qj,Gubser:1998bc} has enjoyed much success in elucidating aspects of strongly-interacting relativistic field theories over the past twelve years. However, there exists a broad array of strongly-coupled systems which admit a field-theoretic description but do not exhibit Lorentz invariance. Cold atomic gases in the so-called unitarity limit provide a prime example of this. The unitarity limit refers to a regime in which the system resides precisely at the crossover point between a gas of Cooper pairs and a Bose-Einstein condensate. This regime arises when the interatomic potential is tuned such that the range of the potential vanishes while the scattering length diverges, thus removing all scales in the problem. The ability to realize this scaling limit experimentally by exploiting the phenomena of Feshbach resonances \cite{OHara} has drawn wide attention to this system and emphasized the need for more progress on the theory front to balance the successes in the laboratory. Recent approaches to this problem that employ field theory techniques directly can be found in \cite{Nishida:2007pj,Nishida:2007mr,Son:2010kq,Nishida:2010tm}.

The unitary Fermi gas is a natural candidate for testing the waters of non-relativistic AdS/CFT as it possesses not just scale invariance but a larger group of symmetries known as the Schr\"{o}dinger group, which can be thought of as the non-relativistic analog of the conformal group. The fact that the Schr\"{o}dinger group is not only a close cousin of conformal symmetry but that it can also be embedded in the conformal group was a key observation in formulating the first proposal for a non-relativistic AdS/CFT correspondence \cite{Son,BM}. This proposal posited that the non-relativistic (Schr\"{o}dinger) correspondence works in much the same way as its relativistic counterpart, but with a particular plane-wave-type geometry (sourced by additional non-trivial background fields) replacing AdS as the dual gravitational spacetime at zero temperature.

The form of the Schr\"{o}dinger spacetime suggests that we should think of the dual Schr\"{o}dinger theory as being embedded within a parent theory living in one higher dimension. If we denote the time coordinate in the parent theory by $t$, then the Schr\"{o}dinger theory is obtained by switching to light cone coordinates $x^{\pm}=y\pm t$, where $y$ is one of the spatial directions, and then projecting the spectrum onto a sector of fixed $x^-$ momentum. The $x^+$ coordinate plays the role of time in the Schr\"{o}dinger theory. If $x^-$ is also compactified, then the $x^-$ momentum assumes a discrete set of values that are naturally interpreted as particle number for the non-relativistic Schr\"{o}dinger theory, and the procedure is referred to as discrete light cone quantization (DLCQ). At zero temperature, there are subtleties involved with this compactification since $x^-$ is lightlike, and one needs to consider large particle number in order to trust the gravity dual \cite{HRR,ABM,MMT}. This idea of embedding the Schr\"{o}dinger theory within a parent theory is closely related to the notion of Bargmann spacetimes \cite{Duvali,Duvalii,Duvaliii}, and it greatly facilitates the formulation of a non-relativistic version of AdS/CFT.

The original conjecture of Refs. \cite{Son,BM} has since been placed on firmer ground by subsequent work which uplifted the dual gravitational background to a solution of IIB supergravity and extended it to a finite temperature black hole solution \cite{HRR,ABM,MMT}. Furthermore, it was shown that at finite temperature, $x^-$ is no longer a lightlike direction, so that subtleties pertaining to its compactification no longer arise. The work of \cite{HRR,ABM,MMT} also clarified that the parent theory related to the Schr\"{o}dinger metric constructed in \cite{Son,BM} is an example of a well known class of non-commutative theories known as dipole theories \cite{Alishahiha:2003ru,Bergman:2000cw,Dasgupta:2000ry}. The dipole theories are obtained by starting from an ordinary commutative field theory and replacing regular multiplication of fields with a non-commutative star product. The particular parent theory described by the Schr\"{o}dinger metric is the dipole theory that results when ${\cal N}{=}4$ super Yang-Mills is deformed in this way, and the corresponding Schr\"{o}dinger theory is just the DLCQ of this. In what follows, we will always refer to these theories simply as the parent and Schr\"{o}dinger theories\footnote{Although Schr\"{o}dinger invariance is broken at finite temperature, we will continue to use the term ``Schr\"{o}dinger theory" even in this case.}, and we will also use these terms to distinguish between their gravity duals, which are also related by DLCQ. These discoveries constitute an important milestone toward extending AdS/CFT to the realm of non-relativistic theories and applying it to improve our understanding of systems like the unitary Fermi gas.

Some of the most important information that AdS/CFT provides about a field theory are its correlation functions. As for theories exhibiting conformal invariance, the two and three-point correlation functions for Schr\"{o}dinger-invariant theories also provide crucial checks of the conjectured correspondence since these correlators are largely determined by the symmetry group alone \cite{Henkel,HU}. In particular, the two-point functions are completely fixed by the symmetry up to an overall constant. Unlike the conformally-invariant case, however, it is not true that the functional form of the three-point functions are fully determined. It was verified in \cite{FM,VW,LH} that AdS/CFT reproduces both two and three-point scalar correlation functions.\footnote{Intriguingly, the authors of \cite{FM} also computed the three-point function using a certain field theory model for a unitary Fermi gas and showed that, in addition, AdS/CFT correctly computes the piece of the three-point function which is not determined by Schr\"{o}dinger symmetry.}

In this paper, we will focus on computing real-time correlation functions for scalar fields in Schr\"{o}dinger field theories. Real-time correlation functions are of particular interest since these in turn yield quantities like conductivity and viscosity. However, real-time correlation functions in AdS/CFT pose more technical challenges relative to Euclidean ones since it is less clear how to formulate basic AdS/CFT recipes in Minkowski signature, where the bulk spacetime tends to be more complicated. Early attempts to deal with this problem yielded a case-by-case treatment \cite{SonStarinets,HerzogSon}, and only in the last two years have more systematic approaches appeared in the literature \cite{Skenderisi,Skenderisii,BVWA}. In the context of Schr\"{o}dinger-invariant theories, the methods of \cite{Skenderisi,Skenderisii} were employed by Hoang and Leigh \cite{LH} to compute time-ordered and Wightman two-point functions at zero temperature. We will show that the approach given in \cite{BVWA} reproduces their findings in a simple way, and we further use it to compute a wide variety of real-time two and three-point correlators at both zero and finite temperature. Our results for the zero-temperature two-point functions are also consistent with the bulk-to-bulk correlators computed in \cite{Blau:2010fh}.

The complete functional form of the zero-temperature two and three-point correlation functions can in fact be computed in real time without having to invoke AdS/CFT or any other method for solving the Schr\"{o}dinger field theory due to the intimate relation between the Schr\"{o}dinger and conformal groups. More specifically, we may obtain correlation functions of the Schr\"{o}dinger theory directly from the correlation functions of a CFT by switching to light cone coordinates and Fourier-transforming with respect to $x^-$, as was originally noticed some time ago in Ref. \cite{HU} and more recently exploited in the context of AdS/CFT in Refs. \cite{FM,VW}. This procedure is essentially equivalent to performing DLCQ, with the role of the Fourier transform being to project onto a sector of fixed $x^-$ momentum. However, as stressed in \cite{HU}, it is important to note that the Fourier-transform trick must be performed keeping $x^-$ noncompact so as to avoid having to construct CFT correlators on a lightlike circle. We will still refer to this procedure as DLCQ since we are free to consider fixed particle number in the resulting expressions, keeping in mind the zero-temperature subtleties mentioned earlier.

We will apply the Fourier-transform technique to compute various zero-temperature correlators in real time. Although the authors of \cite{FM,VW} computed real-time correlators, they did not keep track of the type of correlator (e.g. time-ordered, retarded, etc.), as their primary focus was on testing whether non-relativistic AdS/CFT reproduces Schr\"{o}dinger-invariant expressions. Since the $i\epsilon$ prescriptions which distinguish between the different types of real-time correlators are well known in the case of a relativistic CFT (see \cite{BVWA} for a review), it is a straightforward task to perform DLCQ on these correlators to produce real-time Schr\"{o}dinger correlators, and we will show that the type of correlator is preserved under DLCQ. In addition, we verify that the standard Kallen-Lehmann relations are satisfied by the various real-time Schr\"{o}dinger correlators.

We stress that the CFT correlators that we Fourier transform are not the correlators of the parent theory. In particular, note that the parent dipole theory is not Lorentz-invariant \cite{Alishahiha:2003ru,Bergman:2000cw,Dasgupta:2000ry}. Therefore, the resulting Schr\"{o}dinger-invariant correlators will not contain the same overall constants as those of the DLCQ of the dipole theory. Our aim in applying the Fourier transform to CFT correlators is to ascertain the form of the various real-time Schr\"{o}dinger correlators at zero temperature; this provides us with an important check of the exact Schr\"{o}dinger theory correlators we will compute from non-relativistic AdS/CFT.

We compute the zero-temperature correlators in momentum space as well as position space. DLCQ applied to momentum space correlators simply amounts to performing a rotation in the plane spanned by $\omega$ and $p_y$, the energy and momentum associated with $t$ and $y$. In the case of three-point functions, we express the time-ordered and retarded correlators in a way that is very reminiscent of AdS/CFT, namely in terms of an integral over what are readily interpreted as three bulk-to-boundary propagators in the Schr\"{o}dinger spacetime. This allows us to define in a natural way non-relativistic Feynman and retarded bulk-to-boundary propagators which in turn are easily verified to reproduce the corresponding two-point correlators that we compute directly in field theory. Wightman propagators are then defined in such a way that they reproduce Wightman two-point functions. The exercise of constructing the different real-time bulk-to-boundary propagators and establishing their interrelations not only provides additional consistency checks of non-relativistic AdS/CFT, but also helps set the stage for the computations of thermal correlation functions, which rely heavily on correctly identifying these propagators.

One of the main objectives of this work is to compute real-time two-point and three-point non-relativistic correlation functions for scalar operators dual to minimally coupled scalars at finite temperature. The Schr\"{o}dinger theory at finite temperature also possesses a nonzero chemical potential which breaks Schr\"{o}dinger invariance, meaning that we can no longer receive guidance by applying the Fourier-transform trick to CFT correlators as in the zero-temperature case. With the confidence gained from cross-checking AdS/CFT with field theory at zero temperature, we therefore focus solely on AdS/CFT in the finite temperature case.

The first step is to compute real-time bulk-to-boundary propagators. This is done by starting with the Euclidean version of the metric which describes a black hole in the Schr\"{o}dinger spacetime \cite{HRR,ABM,MMT} and solving the scalar wave equation to obtain the Euclidean bulk-to-boundary propagator in momentum space. This is the Euclidean propagator in the gravity dual of the parent dipole theory. We analytically continue this to a retarded propagator and then perform DLCQ to get the retarded Schr\"{o}dinger bulk-to-boundary propagator. In \cite{BVWA}, it was shown that in the relativistic case, the other real-time thermal propagators can be derived from the retarded one using certain identities. These identities naturally extend to the parent dipole theory, where we can then apply DLCQ to show that similar relations between Schr\"{o}dinger propagators also hold.

Once we have the real-time Schr\"{o}dinger bulk-to-boundary propagators, we can assemble them into real-time two and three-point correlators. This problem was solved in the case of thermal ${\cal N}{=}4$ SYM in \cite{BVWA} by extending Veltman's circling rules \cite{Veltman,'tHooft, kobes} to the gravitational theory on the black hole spacetime, and similar arguments should apply in the present context as well. We compute two-point correlators in this way and confirm that the results satisfy standard Kallen-Lehmann relations. The retarded three-point was worked out explicitly in \cite{BVWA}, and we borrow the result to immediately write down the three-point function for the parent dipole theory. A simple application of DLCQ to the result in turn gives the retarded Schr\"{o}dinger three-point correlator. In addition, we apply the circling rules to construct the time-ordered Schr\"{o}dinger three-point function.

Before proceeding with the calculations, we pause for a moment to briefly discuss the dimensionality of the theories we consider. We keep the number of spatial dimensions (denoted $d$) arbitrary as much as possible. There is certainly no obstacle to doing so in our discussion of DLCQ applied directly to CFT correlators at zero temperature. Furthermore, we may still consider the $d+3$ dimensional Schr\"{o}dinger spacetime as a ``bottom-up" holographic model even though this spacetime can only be uplifted to a solution of IIB supergravity in the case $d=2$. Leaving the dimension arbitrary here also facilitates comparison with our field theory results. We could adopt a similar philosophy toward the Schr\"{o}dinger black hole spacetime, however it is considerably more difficult to solve the wave equation for values of $d$ not equal to 2, so we restrict ourselves to $d=2$ in this case.

We will consider various types of correlators and bulk-to-boundary propagators throughout the paper, and we distinguish between these with different fonts of the letter $g$, as we will now clarify. Relativistic CFT correlators are denoted by $g$, while non-relativistic Schr\"{o}dinger correlators are denoted by $G$. The symbol $\mathfrak g$ specifies bulk-to-boundary propagators in the gravity dual of the parent theory. We also define bulk-to-boundary propagators for the gravity dual of the Schr\"{o}dinger theory, and these we denote by $\cal G$. In addition, we use tildes to denote momentum space functions, the superscript $(3)$ to denote causal three-point correlators, and additional decorations to distinguish between the different types of real-time functions. In particular, the subscripts $F$, $R$, and $A$ denote time-ordered, retarded, and advanced functions respectively. Reverse-time-ordered functions are distinguished from time-ordered functions by an additional bar (e.g. ${\overline G}_F$) which should not be confused with complex conjugation, for which we use the symbol $*$. Wightman two-point functions have superscripts $\pm$, while Wightman three-point functions are specified by a subscript which shows the order of operator insertions. For example, $\widetilde G_{123}$ is a Schr\"{o}dinger Wightman three-point function in momentum space proportional to $\langle O_1O_2O_3\rangle$, where the $O_i$ are arbitrary scalar operators in the Schr\"{o}dinger theory.

The paper is organized as follows. In section \ref{seczerotemptwopt}, we compute all types of scalar real-time Schr\"{o}dinger two-point correlators at zero temperature in both position and momentum space by applying the Fourier-transform trick to relativistic CFT correlators. We also verify that the results satisfy the Kallen-Lehmann relations. In section \ref{secthreept}, we perform similar computations for real-time zero-temperature three-point correlators. We confirm that standard relations among the three-point functions hold, and we also express the results in a way that is very suggestive of an AdS/CFT calculation, allowing us to identify Schr\"{o}dinger ``bulk-to-boundary propagators". In section \ref{seczerotempadscft}, we construct real-time bulk-to-boundary propagators directly in the zero-temperature gravity dual and show that the results are consistent with the answers obtained using the Fourier-transform trick in sections \ref{seczerotemptwopt} and \ref{secthreept}. Finally in section \ref{secfiniteTadscft}, we apply the intuition developed in the zero-temperature case to construct thermal real-time bulk-to-boundary propagators for scalar fields. These are then employed to compute two and three-point scalar correlators for the Schr\"{o}dinger theory at finite temperature. The first four appendices each contain details about the application of the Fourier-transform trick to particular zero-temperature three-point correlators considered in section \ref{secthreept}. In appendix \ref{appendixE}, we construct an explicit example of a massive scalar in the effective 5d gravity theory which arises when a fluctuation in a particular off-diagonal metric component has non-trivial charges on the 5d internal space.

\section{Zero-temperature two-point functions from CFT correlators}\label{seczerotemptwopt}
In this section, we will obtain Schr\"{o}dinger-invariant non-relativistic zero-temperature two-point functions in position space by starting with real-time relativistic conformal two-point functions in light cone coordinates and Fourier-transforming with respect to one of these coordinates. We also obtain momentum space two-point functions by translating the Fourier-transform trick to momentum space, where it becomes a simple redefinition of momenta. We will pay particular attention to how the different types of relativistic real-time functions (e.g. time-ordered, retarded, etc.) map to the different types of non-relativistic functions. We will see that the type of real-time correlator is preserved under the special Fourier transform. We also verify that standard Kallen-Lehmann relations between the various real-time functions are satisfied by the non-relativistic correlators we obtain.

\subsection{Time-ordered and reverse-time-ordered two-point functions in position space}\label{secfeynmantwoptpos}
Consider first the time-ordered relativistic conformal function:
\be
g_F(x^\pm,\vec{x})={-i\over4\pi^2}{1\over(x^+ x^-+x^2+i\epsilon)^\Delta}.
\ee
Here, $x^\pm=y\pm t$, where $t$ is the relativistic time coordinate, and $y$ is one of the spatial coordinates. We use the mostly plus signature for Minkowski space. The coordinate vector $\vec{x}$ represents the remaining $d$ spatial coordinates which coincide with the spatial coordinates of the non-relativistic theory: $\vec{x}=(x_1,x_2,...,x_d)$. Following \cite{HU,VW}, we can obtain a non-relativistic two-point function $G_F$ by performing a Fourier-transform with respect to the $x^-$ direction:
\be
G_F(x^+,\vec{x})={-i\over4\pi^2}\int dx^- {1\over(x^+ x^-+x^2+i\epsilon)^\Delta} e^{-i M x^-}.\label{gffft}
\ee
This integral can be done by introducing a Schwinger parameter:
\be
G_F={(-i)^{\Delta+1}\over4\pi^2\Gamma(\Delta)}\int_0^\infty ds s^{\Delta-1} \int dx^- e^{-[\epsilon-i(x^+x^-+x^2)]s}e^{-iMx^-}.
\ee
The integral on $x^-$ yields a $\delta$-function:
\be
G_F={(-i)^{\Delta+1}\over2\pi\Gamma(\Delta)}\int_0^\infty ds s^{\Delta-1}e^{-[\epsilon-ix^2]s}\delta(M-x^+s).
\ee
The $\delta$-function will evaluate to zero unless $M x^+>0$. Therefore, we find
\be
G_F(x^+,\vec{x})={(-i)^{\Delta+1}\over2\pi\Gamma(\Delta)}|M|^{\Delta-1}\theta(M x^+){e^{iM(x^2+i\epsilon)/x^+}\over|x^+|^{\Delta}}.\label{feynman}
\ee
This has the form of a non-relativistic time-ordered two-point function where the non-relativistic time coordinate is $x^+/2$. Hoang and Leigh \cite{LH} found the same result employing a very different approach \cite{Skenderisi,Skenderisii}. We have retained the $i\epsilon$ in the final expression because it removes a potential singularity at $x^+=0$. It is clear from (\ref{gffft}) that this singularity should not be present.

If we instead start with the reverse-time-ordered relativistic two-point function:
\be
\bar g_F(x^\pm,\vec{x})=-g_F^*={-i\over4\pi^2}{1\over(x^+ x^-+x^2-i\epsilon)^\Delta},
\ee
the same calculation now gives
\be
\overline{G}_F(x^+,\vec{x})={ i^{\Delta-1}\over2\pi\Gamma(\Delta)}|M|^{\Delta-1}\theta(-M x^+){e^{iM(x^2-i\epsilon)/x^+}\over|x^+|^{\Delta}}.
\ee
This has the form of a non-relativistic reverse-time-ordered function.

\subsection{Time-ordered and reverse-time-ordered two-point functions in momentum space}\label{secfeynmantwoptmom}
To make it clear how the Fourier-transform trick translates to momentum space, we first consider a simple example. The Fourier transform of the conformal time-ordered two-point function with $\Delta=1$ and $d=2$ is
\be
\tilde g_F(\omega,p_y,\vec{p})={-i\over4\pi^2}\int dt dyd^2x e^{i\omega t-ip_yy-i\vec{p}\cdot\vec{x}}{1\over -t^2+y^2+x^2+i\epsilon}={1\over \omega^2-p_y^2-p^2+i\epsilon},\label{ffti}
\ee
where $\vec{p}=(p_1,p_2)$ and $p=|\vec{p}|$. Introducing the light cone coordinates $x^\pm$, we can write this as
\bea
\tilde g_F&=&{-i\over8\pi^2}\int dx^+dx^- d^2x e^{i[(\omega-p_y)/2]x^+-i[(\omega+p_y)/2]x^--i\vec{p}\cdot\vec{x}}{1\over x^+x^-+x^2+i\epsilon}\nn\\&=&{1\over \omega^2-p_y^2-p^2+i\epsilon}.\label{fftii}
\eea
We now define the non-relativistic energy $\Omega$ and mass $M$ in terms of the relativistic energy $\omega$ and momentum component $p_y$:
\be
\Omega\equiv {1\over2}(\omega-p_y),\qquad M\equiv {1\over2}(\omega+p_y).\label{momentummap}
\ee
Applying the inverse of this map,
\be
\omega=M+\Omega,\qquad p_y= M-\Omega,\label{momentummapinv}
\ee
we obtain
\bea
\widetilde{G}_F(\Omega,\vec{p})&=&{1\over 4M\Omega -p^2+i\epsilon}={1\over2M}{\theta(M)\over2\Omega-{p^2\over2M}+i\epsilon}+{1\over2M}{\theta(-M)\over2\Omega-{p^2\over2M}-i\epsilon}\nn\\ &=& {1\over2M}{1\over 2\Omega -{p^2\over2M}+i\epsilon M}\label{momentumfeynman}
\eea
This is the non-relativistic time-ordered propagator in momentum space with non-relativistic energy
\be
E_{\mathrm{non-rel.}}=2\Omega.
\ee
It is easy to check that (\ref{momentumfeynman}) is the Fourier transform of (\ref{feynman}) with $\Delta=1$ and $d=2$:\footnote{Our conventions for both relativistic and non-relativistic Fourier transforms are established in this subsection. In particular, note that we define our non-relativistic Fourier transform with an additional overall factor of ${1\over2}$ because the non-relativistic time coordinate is $x^+/2$.}
\be
\widetilde{G}_F(\Omega,\vec{p})={1\over2}\int dx^+ d^2x e^{i\Omega x^+-i\vec{p}\cdot\vec{x}}G_F(x^+,\vec{x}).
\ee

Obviously the same approach can be applied to any other momentum-space function since, from the point of view of the Fourier transform in (\ref{ffti}) and (\ref{fftii}), we are simply redefining momentum-space variables. Starting from the relativistic time-ordered two-point function with arbitrary $\Delta$ and $d$,
\bea
\tilde{g}_F(\omega,p_y,\vec{p})&=&{-i\over4\pi^2}\int dt dyd^2x e^{i\omega t-ip_yy-i\vec{p}\cdot\vec{x}}{1\over ( -t^2+y^2+x^2+i\epsilon)^\Delta}\nn\\&=&{(2i)^{d-2\Delta}\pi^{{d\over2}-1}\Gamma({d\over2}+1-\Delta)\over\Gamma(\Delta)} \left(\omega^2-p_y^2-p^2+i\epsilon\right)^{\Delta-{d\over2}-1},\label{generalfft}
\eea
the map (\ref{momentummapinv}) gives the non-relativistic time-ordered function\footnote{The second line of (\ref{nrfeynmanmomentumtwopt}) can be obtained with the help of the identity $(ab)^\alpha=a^\alpha b^\alpha e^{2\pi i\alpha\eta(a,b)}$, $\eta(a,b)\equiv\theta(-\mathrm{Im}(a))\theta(-\mathrm{Im}(b))\theta(\mathrm{Im}(ab))- \theta(\mathrm{Im}(a))\theta(\mathrm{Im}(b))\theta(-\mathrm{Im}(ab))$.}:
\bea
\widetilde{G}_F(\Omega,\vec{p})&=&{(2i)^{d-2\Delta}\pi^{{d\over2}-1}\Gamma({d\over2}+1-\Delta)\over\Gamma(\Delta)} \left(4M\Omega-p^2+i\epsilon\right)^{\Delta-{d\over2}-1}\nn\\&=&{e^{i\pi[({d\over2}-\Delta)\theta(M)-\theta(-M)]}\pi^{{d\over2}-1}\Gamma({d\over2}+1-\Delta) \over2^{\Delta-{d\over2}+1}\Gamma(\Delta)}|M|^{\Delta-{d\over2}-1} \left[2\Omega-{p^2\over2M}+i\epsilon M\right]^{\Delta-{d\over2}-1}.\nn\\&&\label{nrfeynmanmomentumtwopt}
\eea
The reverse-time-ordered two-point function is just minus the complex conjugate of $\widetilde G_F$:
\bea
&&\widetilde{\overline{G}}_F(\Omega,\vec{p})=-\widetilde{G}_F^*\nn\\&&=-{e^{-i\pi[({d\over2}-\Delta)\theta(M)-\theta(-M)]}\pi^{{d\over2}-1}\Gamma({d\over2}+1-\Delta) \over2^{\Delta-{d\over2}+1}\Gamma(\Delta)}|M|^{\Delta-{d\over2}-1} \left[2\Omega-{p^2\over2M}-i\epsilon M\right]^{\Delta-{d\over2}-1}.\nn\\&&
\eea

\subsection{Wightman two-point functions in position space}\label{secwightmantwoptpos}
Next consider the following relativistic Wightman function:
\be
g^+(x^\pm,\vec{x})={-i\over4\pi^2}{1\over(x^+ x^-+x^2+i\epsilon(x^+-x^-))^\Delta}.\label{rwightman}
\ee
We can facilitate taking the Fourier transform by rewriting this as
\be
g^+={-i\over4\pi^2}{1\over((x^+-i\epsilon)x^-+x^2+i\epsilon (x^+-i\epsilon))^\Delta}.
\ee
Before taking the Fourier transform, it is useful to first factor out a factor of $x^+-i\epsilon$ from the denominator:
\be
G^+(x^+,\vec{x})={-i\over4\pi^2}(x^+-i\epsilon)^{-\Delta}\int dx^-{1\over(x^-+{x^2\over x^+-i\epsilon}+i\epsilon)^\Delta}e^{-iMx^-}.
\ee
One can then use the positivity of $\epsilon$ to introduce a Schwinger parameter as we did in the case of time-ordered functions in the previous subsection. The $x^-$ integral then gives a $\delta$-function:
\be
G^+={(-i)^{\Delta+1} (x^+-i\epsilon)^{-\Delta}\over 2\pi\Gamma(\Delta)}\int_0^\infty ds s^{\Delta-1}e^{-[\epsilon-ix^2/(x^+-i\epsilon)]s}\delta(M-s).
\ee
Since the argument of the $\delta$-function does not contain $x^+$, a step function involving $x^+$ does not arise in this case. The result is then
\be
G^+(x^+,\vec{x})={(-i)^{\Delta+1}\over2\pi\Gamma(\Delta)}M^{\Delta-1}\theta(M){e^{iMx^2/(x^+-i\epsilon)}\over(x^+-i\epsilon)^{\Delta}}.\label{nrwightmanplus}
\ee
This agrees with the result found in \cite{LH} using the methods of \cite{Skenderisi,Skenderisii}. The role of the $i\epsilon$ in the exponent is to ensure that $G^+$ vanishes at $x^+=0$, while the $i\epsilon$ in the denominator controls the location of the branch cut for non-integer $\Delta$:\footnote{Throughout this paper, we adopt the convention that $(-1)^\Delta=e^{i\pi\Delta}$ if no $i\epsilon$ prescription is given explicitly.}
\be
(x^+-i\epsilon)^{-\Delta}=\theta(x^+)|x^+|^{-\Delta}+\theta(-x^+)e^{i\pi\Delta}|x^+|^{-\Delta}.\label{iepsprescription}
\ee

If we instead consider starting with the other relativistic Wightman function,
\be
g^-(x^\pm,\vec{x})={-i\over4\pi^2}{1\over(x^+x^-+x^2-i\epsilon (x^+-x^-))^\Delta},
\ee
the same steps lead to an $s$-integration as above, but with $\delta(M+s)$ instead of $\delta(M-s)$ appearing in the integrand, yielding
\be
G^-(x^+,\vec{x})={ i^{\Delta-1}\over2\pi\Gamma(\Delta)}|M|^{\Delta-1}\theta(-M){e^{iMx^2/(x^++i\epsilon)}\over(x^++i\epsilon)^{\Delta}}.\label{nrwightmanminus}
\ee
Using (\ref{feynman}), (\ref{nrwightmanplus}), and (\ref{nrwightmanminus}) it is easy to check that the relations
\bea
G_F(x^+,\vec{x})&=&\theta(x^+)G^+(x^+,\vec{x})+\theta(-x^+)G^-(x^+,\vec{x}),\nn\\\overline G_F(x^+,\vec{x})&=&\theta(-x^+)G^+(x^+,\vec{x})+\theta(x^+) G^-(x^+,\vec{x}),\label{GfGpGmrelation}
\eea
are satisfied.\footnote{When $x^+\ne0$, the $i\epsilon$ in the exponents can be discarded relative to $x^+$. When $x^+=0$, $G_F$, $G^+$, and $G^-$ each vanish identically, and (\ref{GfGpGmrelation}) is satisfied trivially.}

\subsection{Wightman two-point functions in momentum space}\label{secwightmantwoptmom}
Now consider the relativistic Wightman functions with $\Delta=1$ and $d=2$ in momentum space:
\be
\tilde{g}^\pm(\omega,p_y,\vec{p}) = -2\pi i\theta(\pm \omega)\delta(\omega^2-p_y^2-p^2).
\ee
Under (\ref{momentummapinv}), these transform to
\be
\widetilde{G}^\pm(\Omega,\vec{p})=-2\pi i\theta(\pm(\Omega+M))\delta(4M\Omega-p^2).
\ee
For $M>0$ ($M<0$), the $\delta$-function has support at positive (negative) values of $\Omega$, so that
\be
\widetilde{G}^+(\Omega,\vec{p})=-2\pi i\theta(M)\delta(4M\Omega-p^2),
\ee
which is the Fourier transform of (\ref{nrwightmanplus}) for $\Delta=1$, $d=2$, and
\be
\widetilde{G}^-(\Omega,\vec{p})=-2\pi i\theta(-M)\delta(4M\Omega-p^2).
\ee

To obtain the non-relativistic Wightman functions for general $\Delta$ and $d$, we start with the Fourier transform of (\ref{rwightman}):
\be
\tilde{g}^+(\omega,p_y,\vec{p})=-i{2^{d+1-2\Delta}\pi^{{d\over2}}\over\Gamma(\Delta)\Gamma(\Delta-{d\over2})}\theta(\omega-\sqrt{p_y^2+p^2})(\omega^2-p_y^2-p^2)^{\Delta-{d\over2}-1}.
\ee
The map (\ref{momentummapinv}) then gives
\be
\widetilde{G}^+(\Omega,\vec{p})=-i{2^{d+1-2\Delta}\pi^{{d\over2}}\over\Gamma(\Delta)\Gamma(\Delta-{d\over2})}\theta(M)\theta(4M\Omega-p^2)(4M\Omega-p^2)^{\Delta-{d\over2}-1}.
\ee
We have used that
\be
\theta(\omega-\sqrt{p_y^2+p^2})=\theta(\omega+p_y)\theta(\omega^2-p_y^2-p^2)
\ee
to rewrite the argument of the step function before applying the map. The second relativistic Wightman function has the form
\be
\tilde{g}^-(\omega,p_y,\vec{p})=-i{2^{d+1-2\Delta}\pi^{{d\over2}}\over\Gamma(\Delta)\Gamma(-\Delta-{d\over2})}\theta(-\omega-\sqrt{p_y^2+p^2})(\omega^2-p_y^2-p^2)^{\Delta-{d\over2}-1}.
\ee
Now observing that
\be
\theta(-\omega-\sqrt{p_y^2+p^2})=\theta(\omega^2-p_y^2-p^2)\theta(-\omega-p_y),
\ee
we see that the map (\ref{momentummapinv}) leads to
\be
\widetilde{G}^-(\Omega,\vec{p})=-i{2^{d+1-2\Delta}\pi^{{d\over2}}\over\Gamma(\Delta)\Gamma(\Delta-{d\over2})}\theta(-M)\theta(4M\Omega-p^2)(4M\Omega-p^2)^{\Delta-{d\over2}-1}.
\ee
So far, we have seen that the relativistic (reverse-)time-ordered two-point function transforms into a non-relativistic (reverse-)time-ordered two-point function, and the relativistic Wightman functions map to non-relativistic Wightman functions. It remains for us to check what happens to relativistic retarded and advanced functions under this mapping.

\subsection{Retarded and advanced two-point functions in position space}\label{secretardedtwoptpos}
The relativistic conformal retarded two-point correlator can be written in terms of Wightman functions:
\be
g_R(t,y,\vec{x})=\theta(t)\left[g^+(t,y,\vec{x})-g^-(t,y,\vec{x})\right].
\ee
In terms of light cone coordinates, we therefore have
\bea
g_R(x^\pm,\vec{x})&=&{-i\over4\pi^2}\theta(x^+-x^-)\bigg[{1\over((x^+-i\epsilon)x^-+x^2+i\epsilon (x^+-i\epsilon)))^\Delta}\nn\\&-&{1\over((x^++i\epsilon)x^-+x^2-i\epsilon (x^++i\epsilon))^\Delta}\bigg].\label{rretarded}
\eea
Factoring out the $(x^+ {\mp} i\epsilon)$ in the denominators, assuming $x^+\ne0$, and Fourier-transforming leads to
\bea
G_R(x^+,\vec{x})&=&{-i\over4\pi^2}(x^+-i\epsilon)^{-\Delta}\int_{-\infty}^{x^+} dx^-{e^{-iMx^-}\over(x^-+x^2/x^++i\epsilon)^\Delta}\nn\\&+&{i\over4\pi^2}(x^++i\epsilon)^\Delta
\int_{-\infty}^{x^+}dx^-{e^{-iMx^-}\over(x^-+x^2/x^+-i\epsilon)^\Delta}.\label{GRintegrals}
\eea
Notice that we have chosen to write $x^2/x^+$ instead of $x^2/(x^+\pm i\epsilon)$ in the denominators. Since we are assuming $x^+\ne0$, we may rewrite $x^2/(x^+\pm i\epsilon)=x^2/x^+\mp i\epsilon x^2/(x^+)^2$, and the term $i\epsilon x^2/(x^+)^2$ can be eliminated by rescaling the $\epsilon$ in the denominators of the integrands. The $i\epsilon$ in the factors $(x^+\pm i\epsilon)^{-\Delta}$ appearing outside the integrals cannot be neglected even though $x^+\ne0$ because the $i\epsilon$ in these factors control branch cuts, as we discussed in the previous subsection on Wightman functions.

Changing the integration variable to $u\equiv x^-+x^2/x^+$, we have
\bea
G_R&=&{-i\over4\pi^2}(x^+-i\epsilon)^{-\Delta}e^{iMx^2/x^+}\int_{-\infty}^{x^2/x^++x^+} du{e^{-iMu}\over(u+i\epsilon)^\Delta}\nn\\&+&{i\over4\pi^2}(x^++i\epsilon)^{-\Delta}\int_{-\infty}^{x^2/x^++x^+}du{e^{-iMu}\over(u-i\epsilon)^\Delta}.
\eea
The integrals can be evaluated using
\be
\lim_{\epsilon\to0}\int_{-\infty}^\alpha du{e^{-iM u}\over(u\pm i\epsilon)^\Delta}=\pm {2\pi(-i)^\Delta\over\Gamma(\Delta)}M^{\Delta-1}\theta(\alpha)\theta(\pm M),
\ee
and we find
\be
G_R(x^+,\vec{x})={(-i)^{\Delta+1}\over2\pi\Gamma(\Delta)}|M|^{\Delta-1}e^{i\pi(\Delta-1)\theta(-M)}\theta(x^+)|x^+|^{-\Delta}e^{iMx^2/x^+},\quad x^+\ne0.\label{retardedtwoptposition}
\ee
This has the form of a non-relativistic retarded two-point function. This expression is valid for $x^+\ne0$; when $x^+=0$, it is easy to check that $G_R=0$. (This can be done by setting $x^+=0$ in (\ref{rretarded}) and Fourier-transforming.) An analogous calculation reveals that if we start with the relativistic advanced function, we obtain the non-relativistic advanced function:
\be
G_A(x^+,\vec{x})={(-i)^{\Delta+1}\over2\pi\Gamma(\Delta)}|M|^{\Delta-1}e^{i\pi(\Delta-1)\theta(M)}\theta(-x^+)|x^+|^{-\Delta}e^{iMx^2/x^+},\quad x^+\ne0.\label{advancedtwoptposition}
\ee
A shortcut to obtaining this result is to write down the analog of (\ref{GRintegrals}) for $G_A$ and to notice that $G_A$ can be obtained from $G_R$ by sending $x^+\to-x^+$ and $M\to-M$ in the final answer. It is easy to check using (\ref{nrwightmanplus}) and (\ref{nrwightmanminus}) that standard relations between retarded/advanced and Wightman functions:
\be
G_R=\theta(x^+)[G^+-G^-],\qquad G_A=\theta(-x^+)[G^--G^+],
\ee
hold for the non-relativistic two-point functions we have obtained.

We conclude that the type of real-time two-point functions is preserved under the mapping which takes us from relativistic to non-relativistic functions.

\subsection{Retarded and advanced two-point functions in momentum space}\label{secretardedtwoptmom}
The Fourier transform of the relativistic retarded two-point function (\ref{rretarded}) with $\Delta=1$ and $d=2$ is given by
\be
\tilde{g}_R(\omega,p_y,\vec{p})=\theta(\omega)\tilde{g}_F+\theta(-\omega)\tilde{g}_F^*={1\over \omega^2-p_y^2-p^2+i\epsilon\omega}.
\ee
Applying (\ref{momentummapinv}), we obtain
\bea
\widetilde{G}_R&=&\theta(\Omega+M)\widetilde{G}_F+\theta(-\Omega-M)\widetilde{G}^*_F={\theta(\Omega+M)\over4M\Omega-p^2+ i\epsilon}+{\theta(-\Omega-M)\over4M\Omega-p^2-i\epsilon}\nn\\&=&{1\over 4M\Omega-p^2+i\epsilon(\Omega+M)}.\label{retardedtwoptfromfeynman}
\eea
This function has a pole at
\be
\Omega_{p}={p^2\over4M}-i\epsilon\left({1\over4}+{p^2\over16M^2}\right).
\ee
Since this pole lies in the lower half $\Omega$-plane regardless of the value of $M$ or $p^2$, we see that this function has the analytic properties expected of a retarded Green's function. We may therefore redefine $\epsilon$ to absorb the positive coefficient which multiplies it in $\Omega_{p}$:
\be
\widetilde{G}_R(\Omega,\vec{p})={1\over2M(2\Omega-{p^2\over2M}+i\epsilon)}.\label{retardedtwoptmomentum}
\ee
The advanced function is the conjugate of this:
\be
\widetilde{G}_A=\widetilde{G}_R^*.
\ee
Comparing these results with (\ref{momentumfeynman}), we see that $\widetilde G_F$, $\widetilde G_R$ and $\widetilde G_A$ obey the following relation:
\be
\widetilde{G}_F(\Omega,\vec{p})=\theta(M)\widetilde G_R(\Omega,\vec{p})+\theta(-M)\widetilde G_A(\Omega,\vec{p}).\label{GFGRGArelation}
\ee
On the other hand, we can obtain what appears to be a slightly different relation directly from (\ref{retardedtwoptfromfeynman}):
\be
\widetilde{G}_F(\Omega,\vec{p})=\theta(\Omega+M)\widetilde G_R(\Omega,\vec{p})+\theta(-\Omega-M)\widetilde G_A(\Omega,\vec{p}).\label{GFGRGArelationii}
\ee
At first glance, it may seem surprising that these two relations hold simultaneously. However, their compatibility has a simple origin in the form of the imaginary part of the retarded two-point correlator. Using (\ref{retardedtwoptfromfeynman}), we can write this as $\mathrm{Im}\widetilde G_R\sim \mathrm{sgn}(\Omega+M)\delta(4M\Omega-p^2)$. Note that only the imaginary part is affected by the presence of the step functions in (\ref{GFGRGArelation}) and (\ref{GFGRGArelationii}). Since the $\delta$-function requires that $\Omega$ and $M$ have the same sign, this is equivalent to $\mathrm{Im}\widetilde G_R\sim \mathrm{sgn}(M)\delta(4M\Omega-p^2)$, which agrees with (\ref{retardedtwoptmomentum}). Also notice that we can write the relation in a third way:
\be
\widetilde{G}_F(\Omega,\vec{p})=\theta(\Omega)\widetilde G_R(\Omega,\vec{p})+\theta(-\Omega)\widetilde G_A(\Omega,\vec{p}).\label{GFGRGArelationiii}
\ee
This form of the relation makes it apparent that there is no chemical potential at zero temperature. We will now see that the relations (\ref{GFGRGArelation}), (\ref{GFGRGArelationii}), and (\ref{GFGRGArelationiii}) hold for general values of $\Delta$ and $d$.

The Fourier transform of (\ref{rretarded}) for general $\Delta$ and $d$ is
\be
\tilde{g}_R(\omega,p_y,\vec{p})=\theta(\omega)\tilde{g}_F+\theta(-\omega)\tilde{g}_F^*,
\ee
where $\tilde{g}_F$ was given in (\ref{generalfft}). This yields
\bea
\widetilde{G}_R(\Omega,\vec{p})&=&\theta(\Omega+M)\widetilde{G}_F+\theta(-\Omega-M)\widetilde{G}^*_F\nn\\&=& {(2i\mathrm{sgn}(\Omega+M))^{d-2\Delta}\pi^{{d\over2}-1}\Gamma({d\over2}+1-\Delta)\over\Gamma(\Delta)} \left(4M\Omega-p^2+i\epsilon (\Omega+M)\right)^{\Delta-{d\over2}-1}\nn\\&=&{e^{i\pi[({d\over2}-\Delta)\theta(M)-\theta(-M)]}\pi^{{d\over2}-1}\Gamma({d\over2}+1-\Delta) \over2^{\Delta-{d\over2}+1}\Gamma(\Delta)}|M|^{\Delta-{d\over2}-1} \left[2\Omega-{p^2\over2M}+i\epsilon \right]^{\Delta-{d\over2}-1}.\nn\\&&\label{nrretardedmomentumtwopt}
\eea
Similar reasoning leads to the following advanced two-point function:
\bea
\widetilde{G}_A(\Omega,\vec{p})&=&\theta(\Omega+M)\widetilde{G}^*_F+\theta(-\Omega-M)\widetilde{G}_F\nn\\&=& {(-2i\mathrm{sgn}(\Omega+M))^{d-2\Delta}\pi^{{d\over2}-1}\Gamma({d\over2}+1-\Delta)\over\Gamma(\Delta)} \left(4M\Omega-p^2-i\epsilon (\Omega+M)\right)^{\Delta-{d\over2}-1}\nn\\&=&{e^{-i\pi[({d\over2}-\Delta)\theta(M)-\theta(-M)]}\pi^{{d\over2}-1}\Gamma({d\over2}+1-\Delta) \over2^{\Delta-{d\over2}+1}\Gamma(\Delta)}|M|^{\Delta-{d\over2}-1} \left[2\Omega-{p^2\over2M}-i\epsilon \right]^{\Delta-{d\over2}-1}.\nn\\&&
\eea
It is not difficult to check that the relations (\ref{GFGRGArelation}), (\ref{GFGRGArelationii}), and (\ref{GFGRGArelationiii}) hold for general $\Delta$ and $d$. Furthermore, the relation (\ref{GFGRGArelation}) provides a useful check of the consistency of our results since it must also hold in position space, and one can readily verify that (\ref{feynman}), (\ref{retardedtwoptposition}), and (\ref{advancedtwoptposition}) satisfy (\ref{GFGRGArelation}). We will see later on in section \ref{secfiniteTadscft} that the second relation, (\ref{GFGRGArelationii}), generalizes to a relation that is also valid at finite temperature.

\section{Zero-temperature three-point functions from CFT correlators}\label{secthreept}
In this section, we apply the Fourier-transform trick to relativistic three-point functions to obtain Schr\"{o}dinger-invariant non-relativistic three-point functions in position space. We do this for time-ordered and Wightman functions and show that the usual identity relating these types of correlators is satisfied. In position space, retarded and advanced functions can be expressed in terms of the Wightman functions \cite{ls}.

We compute non-relativistic time-ordered and retarded/advanced three-point functions in momentum space as well. This can be done either by computing the Fourier transforms of the non-relativistic position space three-point functions, or by starting with relativistic conformal three-point functions in momentum space and performing an appropriate redefinition of momenta as we did for two-point functions in section \ref{seczerotemptwopt}. We will apply the former approach to obtain time-ordered functions and the latter approach for retarded/advanced functions. Wightman three-point functions are quite unwieldy in momentum space, so we do not include them here.

The final answers we obtain for the momentum space three-point functions are of the form one would expect from an AdS/CFT calculation. In particular, we express the results as integrals of products of three functions which are naturally interpreted as bulk-to-boundary propagators for scalar fields in non-relativistic AdS/CFT \cite{Son,BM}. In section \ref{seczerotempadscft}, we reproduce these bulk-to-boundary propagators by applying DLCQ to propagators in the gravity dual of the dipole theory and show that they give rise to two-point functions which are consistent with the non-relativistic correlators we computed in section \ref{seczerotemptwopt}.

\subsection{Time-ordered three-point functions in position space}\label{secfeynmanthreeptpos}
The relativistic time-ordered three-point function in position space has the form
\bea
g^{(3)}_F&=&-\langle0|{\cal T}O_1(x_1)O_2(x_2)O_3(x_3)|0\rangle=-{1\over[x_{23}^+x_{23}^-+x_{23}^2+i\epsilon]^{\delta_1}}\nn\\&\times&{1\over[x_{13}^+x_{13}^-+x_{13}^2+i\epsilon]^{\delta_2}} {1\over[x_{12}^+x_{12}^-+x_{12}^2+i\epsilon]^{\delta_3}}.\label{relativisticfeynmanthreeptpos}
\eea
We have introduced the notation $x_{ij}=x_i-x_j$. The constants $\delta_i$ are related to the conformal dimensions $\Delta_i$ of the scalar fields:
\be
\delta_1\equiv {1\over2}(\Delta_2+\Delta_3-\Delta_1),\quad \delta_2\equiv {1\over2}(\Delta_1+\Delta_3-\Delta_2), \quad \delta_3\equiv {1\over2}(\Delta_1+\Delta_2-\Delta_3).\label{defofdelta}
\ee
To obtain the non-relativistic correlator $G_F^{(3)}$, we must Fourier-transform with respect to the $x_i^-$. The integrals can be computed with the help of Schwinger parameters, and the details can be found in appendix \ref{appendixA}. The result is
\bea
G^{(3)}_F&=&-{(2\pi)^3(-i)^{\delta_1+\delta_2+\delta_3}\delta(\sum_iM_i)\over\Gamma(\delta_1)\Gamma(\delta_2)\Gamma(\delta_3)}|x_{23}^+|^{-\delta_1}|x_{13}^+|^{-\delta_2} e^{iM_1(x_{13}^2+i\epsilon)/x_{13}^+}e^{iM_2(x_{23}^2+i\epsilon)/x_{23}^+}I(v_{12}),\nn\\&&\label{feynmanthreeptposition}
\eea
where
\be
v_{12}\equiv {x_{12}^2+i\epsilon\over x_{12}^+}+{x_{23}^2+i\epsilon\over x_{23}^+}-{x_{13}^2+i\epsilon\over x_{13}^+},\label{vonetwodef}
\ee
and
\bea
I&=&\theta(x_{12}^+)\theta(x_{13}^+)\theta(x_{23}^+)I_{+++}+\theta(x_{12}^+)\theta(x_{13}^+)\theta(x_{32}^+)I_{++-}+\theta(x_{12}^+)\theta(x_{31}^+)\theta(x_{32}^+)I_{+--}\nn\\&+& \theta(x_{21}^+)\theta(x_{13}^+)\theta(x_{23}^+)I_{-++}+\theta(x_{21}^+)\theta(x_{31}^+)\theta(x_{23}^+)I_{--+}+\theta(x_{21}^+)\theta(x_{31}^+)\theta(x_{32}^+)I_{---}.\nn\\&&\label{defofI}
\eea
The functions $I_{ijk}$ appearing in this expression are given by
\bea
I_{+++}&=&|x_{12}^+|^{-\delta_3}\theta(M_1)\bigg\{\theta(M_2)M_1^{\delta_2+\delta_3-1}M_2^{\delta_1-1}B(\delta_3,\delta_2)\Phi_1(\delta_3, 1-\delta_1,\delta_2+\delta_3,-{M_1\over M_2},iv_{12}M_1)\nn\\&+&\theta(-M_2)\theta(M_1-|M_2|)(M_1-|M_2|)^{\delta_1+\delta_2-1}|M_2|^{\delta_3-1}e^{iv_{12}|M_2|}\nn\\&\times& B(\delta_1,\delta_2)\Phi_1(\delta_1,1-\delta_3,\delta_1+\delta_2,{|M_2|-M_1\over|M_2|},iv_{12}(M_1-|M_2|))\bigg\},\label{Ippp}
\eea
\bea
I_{++-}&=&|x_{12}^+|^{-\delta_3}\int_0^\infty dzz^{\delta_3-1}|M_2+z|^{\delta_1-1}|M_1-z|^{\delta_2-1}e^{iv_{12}z}\theta(M_1-z)\theta(-M_2-z)\nn\\&=&|x_{12}^+|^{-\delta_3}\theta(M_1)\theta(-M_2)\bigg\{\theta(M_1-|M_2|)M_1^{\delta_2-1}|M_2|^{\delta_1+\delta_3-1}B(\delta_3,\delta_1)\nn\\&\times&\Phi_1(\delta_3, 1-\delta_2,\delta_1+\delta_3,{|M_2|\over M_1},iv_{12}|M_2|)+\theta(|M_2|-M_1)M_1^{\delta_2+\delta_3-1}|M_2|^{\delta_1-1}\nn\\&\times&B(\delta_3,\delta_2)\Phi_1(\delta_3,1-\delta_1,\delta_2+\delta_3,{M_1\over|M_2|},iv_{12}M_1)\bigg\},\label{Ippm}
\eea
\bea
I_{+--}&=&|x_{12}^+|^{-\delta_3}\int_0^\infty dzz^{\delta_3-1}|M_2-z|^{\delta_1-1}|M_1+z|^{\delta_2-1}e^{-iv_{12}z}\theta(M_1+z)\theta(M_2-z)\nn\\&=&|x_{12}^+|^{-\delta_3}\theta(-M_2)\bigg\{\theta(M_1)\theta(|M_2|-M_1)e^{iv_{12}M_1}M_1^{\delta_3-1}(|M_2|-M_1)^{\delta_1+\delta_2-1}\nn\\&\times& B(\delta_2,\delta_1)\Phi_1(\delta_2,1-\delta_3,\delta_1+\delta_2,{M_1-|M_2|\over M_1},iv_{12}(|M_2|-M_1))\nn\\&+&\theta(-M_1)|M_1|^{\delta_2-1}|M_2|^{\delta_1+\delta_3-1} B(\delta_3,\delta_1)\Phi_1(\delta_3,1-\delta_2,\delta_1+\delta_3,-{M_2\over M_1},iv_{12}|M_2|)\bigg\}.\nn\\&&\label{Ipmm}
\eea
The function $B(x,y)$ is the Euler beta function, while $\Phi_1$ is a hypergeometric function of two variables. These functions are defined in (\ref{defeulerbeta}) and (\ref{defhypgeom}) respectively. The remaining three $I_{ijk}$ appearing in (\ref{defofI}) can be obtained from the ones quoted above using the relations
\bea
I_{-++}(\delta_1,\delta_2,\delta_3,M_1,M_2,v_{12})&=&I_{+++}(\delta_2,\delta_1,\delta_3,M_2,M_1,-v_{12}),\nn\\
I_{--+}(\delta_1,\delta_2,\delta_3,M_1,M_2,v_{12})&=&I_{++-}(\delta_2,\delta_1,\delta_3,M_2,M_1,-v_{12}),\nn\\
I_{---}(\delta_1,\delta_2,\delta_3,M_1,M_2,v_{12})&=&I_{+--}(\delta_2,\delta_1,\delta_3,M_2,M_1,-v_{12}).\label{Irelations}
\eea
We will see in section \ref{secwightmanthreeptpos} that the $I_{ijk}$ are proportional to the Wightman functions $G_{ijk}$ in precisely the way one would expect from the standard identity
\bea
G_F^{(3)}&=&\theta(x_{12}^+)\theta(x_{13}^+)\theta(x_{23}^+)G_{123}+\theta(x_{12}^+)\theta(x_{13}^+)\theta(x_{32}^+)G_{132}+\theta(x_{12}^+)\theta(x_{31}^+)\theta(x_{32}^+)G_{312}\nn\\&+& \theta(x_{21}^+)\theta(x_{13}^+)\theta(x_{23}^+)G_{213}+\theta(x_{21}^+)\theta(x_{31}^+)\theta(x_{23}^+)G_{231}+\theta(x_{21}^+)\theta(x_{31}^+)\theta(x_{32}^+)G_{321}.\nn\\&& \label{feynmanwightmanthreeptrelation}
\eea

\subsection{Time-ordered three-point functions in momentum space}\label{secfeynmanthreeptmom}
We next consider time-ordered three-point functions in momentum space. We could compute these by starting with conformally-invariant time-ordered three-point functions in momentum space and performing a redefinition of momenta, but we choose instead to Fourier-transform the CFT correlator in position space, eqn. (\ref{relativisticfeynmanthreeptpos}). As in previous sections, the Fourier transform is computed by introducing Schwinger parameters; the details can be found in appendix \ref{appendixB}. The result is
\bea
\widetilde{G}^{(3)}_F&=&{2^{5d/2+8-2\sum_i\delta_i}\pi^{2d+4}\over\Gamma(\delta_1)\Gamma(\delta_2)\Gamma(\delta_3)\Gamma(\sum_i\delta_i-{d\over2}-1)}\delta(\sum_i\Omega_i)\delta(\sum_iM_i)\delta^d(\sum_i\vec{p}_i) \nn\\&\times&\int_0^\infty dzz^{d\over2}\prod_j(-4M_j\Omega_j+p_j^2-i\epsilon)^{\nu_j/2}\mathrm{K}_{\nu_j}\left(z\sqrt{-4M_j\Omega_j+p_j^2-i\epsilon}\right),\label{feynmanthreeptmomentum}
\eea
where we have defined
\be
\nu_j\equiv \sum_{i\ne j}\delta_i - {d\over2}-1=\Delta_j-{d\over2}-1.\label{defofnu}
\ee
Our result for $\widetilde G^{(3)}_F$ has precisely the structure one would expect to see arise out of an AdS/CFT calculation. This observation suggests that we can identify a non-relativistic time-ordered ``bulk-to-boundary propagator":
\be
\widetilde{\cal G}_F(\Omega,\vec{p};M,\Delta)\sim z^{{d\over2}+1}(-4M\Omega+p^2-i\epsilon)^{\nu/2}\mathrm{K}_{\nu}\left(z\sqrt{-4M\Omega+p^2-i\epsilon}\right).\label{feynmanprop}
\ee
$z$ can be interpreted as a radial coordinate in (a modified version of) AdS. We have put in by hand the factor of $z^{{d\over2}+1}$ appearing in $\widetilde{\cal G}_F$; this will be justified later by invoking the proposed AdS/CFT dual \cite{Son,BM}, which we will review and make use of in section \ref{seczerotempadscft}. Note that after separating out a factor of $z^{{d\over2}+1}$ for each propagator, there remains the factor $z^{-d-3}$, which is precisely the measure factor of the dual gravity spacetime.

We can work out the position-space analog of (\ref{feynmanthreeptmomentum}) by computing its Fourier transform, with the result
\bea
G_F^{(3)}(x^+_i,\vec{x}_i;M_i)&=&-{2^{4-{1\over2}\sum_i\nu_i}\pi^{6-{d\over2}}\Gamma(\Delta_1)\Gamma(\Delta_2)\Gamma(\Delta_3)\over\Gamma(\delta_1)\Gamma(\delta_2)\Gamma(\delta_3)\Gamma(\sum_i \delta_i-{d\over2}-1)}\delta(\sum_iM_i) \int_0^\infty dz \int dx^+d^dx z^{-d-3}\nn\\&\times&{\cal G}_F(x_1-x,z;M_1,\Delta_1){\cal G}_F(x_2-x,z;M_2,\Delta_2){\cal G}_F(x_3-x,z;M_3,\Delta_3),\nn\\&&
\eea
where
\be
{\cal G}_F(x,z;M,\Delta)={(-i)^{\Delta+1}\over2\pi\Gamma(\Delta)}|M|^{\Delta-1}z^\Delta\theta(Mx^+) |x^+|^{-\Delta}e^{iM(x^2+z^2)/x^+}
\ee
can be interpreted as a Schr\"{o}dinger Feynman bulk-to-boundary propagator in position space in the sense that it is the Fourier transform of (\ref{feynmanprop}). A similar expression was obtained by Volovich and Wen \cite{VW}.

\subsection{Wightman three-point functions in position space}\label{secwightmanthreeptpos}
A conformally-invariant Wightman function in position space can be expressed in the form
\bea
g_{123}(x_1,x_2,x_3)&\equiv&-\langle0| O_1(x_1)O_2(x_2)O_3(x_3)|0\rangle=-{1\over [x_{23}^+x_{23}^-+x_{23}^2+i\epsilon(x_{23}^+-x_{23}^-)]^{\delta_1}}\nn\\&\times&{1\over [x_{13}^+x_{13}^-+x_{13}^2+i\epsilon(x_{13}^+-x_{13}^-)]^{\delta_2}}{1\over [x_{12}^+x_{12}^-+x_{12}^2+i\epsilon(x_{12}^+-x_{12}^-)]^{\delta_3}}.\nn\\&&
\eea
The $i\epsilon$ prescription is such that the time coordinate $t_i$ of operator $O_i$ is greater by $-i\epsilon$ than the time coordinate of any operator to the right of $O_i$ \cite{wightman,Hofman}. We compute the associated non-relativistic three-point correlator by Fourier-transforming with respect to the $x_i^-$. This is done in appendix \ref{appendixC}, and one finds
\bea
G_{123}&=&-{(2\pi)^3(-i)^{\delta_1+\delta_2+\delta_3}\delta(\sum_iM_i)\over \Gamma(\delta_1)\Gamma(\delta_2)\Gamma(\delta_3)}(x_{23}^+-i\epsilon)^{-\delta_1}(x_{13}^+-i\epsilon)^{-\delta_2}(x_{12}^+-i\epsilon)^{-\delta_3}\nn\\&\times& e^{iM_1x_{13}^2{x_{13}^+\over(x_{13}^+)^2+\epsilon^2}}e^{iM_2x_{23}^2{x_{23}^+\over(x_{23}^+)^2+\epsilon^2}} \theta(M_1)\bigg\{\theta(M_2)M_1^{\delta_2+\delta_3-1}M_2^{\delta_1-1}B(\delta_3,\delta_2)\nn\\&\times&\Phi_1(\delta_3, 1-\delta_1,\delta_2+\delta_3,-{M_1\over M_2},iv_{12}M_1)+\theta(-M_2)\theta(M_1-|M_2|)e^{iv_{12}|M_2|}|M_2|^{\delta_3-1}\nn\\&\times&(M_1-|M_2|)^{\delta_1+\delta_2-1} B(\delta_1,\delta_2)\Phi_1(\delta_1,1-\delta_3,\delta_1+\delta_2,{|M_2|-M_1\over|M_2|},iv_{12}(M_1-|M_2|))\bigg\},\nn\\&&\label{wightmanthreeptposition}
\eea
with
\be
v_{12}\equiv {x_{12}^2\over x_{12}^+-i\epsilon}+{x_{23}^2\over x_{23}^+-i\epsilon}-{x_{13}^2\over x_{13}^+-i\epsilon}.
\ee
The functions $B$ and $\Phi_1$ are defined in appendix \ref{appendixA}. This expression is consistent with (\ref{feynmanwightmanthreeptrelation}) as can be seen by noting that when $x^+$ is positive, $(x^+-i\epsilon)^{-\delta}=|x^+|^{-\delta}$. The other Wightman three-point functions can of course be obtained by permuting the labels 1,2,3. The relations given in (\ref{Irelations}) can then simply be understood as particular examples of these permutations in which the labels 1 and 2 are swapped. These permutations are particularly simple because the coefficient in front of $I$ in (\ref{feynmanthreeptposition}) is symmetric with respect to the swapping of 1 and 2. Permutations of either 1 or 2 with 3 are less trivial because this coefficient does not exhibit symmetry under these swappings. These cases can be checked by using an integral expression for $I$ given in eqn. (\ref{Iintegral}).

\subsection{Retarded and advanced three-point functions in momentum space}\label{secretardedthreeptmom}
For non-relativistic retarded and advanced three-point functions, we will apply the approach used in section \ref{seczerotemptwopt} for obtaining non-relativistic momentum space two-point functions directly from relativistic functions in momentum space. There, we saw that we can simply start with the relativistic function and rewrite the relativistic momenta in terms of non-relativistic momenta. The relativistic retarded three-point function was computed in \cite{BVWA}\footnote{In ref. \cite{BVWA}, the retarded function was computed in 3+1 dimensions in the case where all three scalar fields have the same conformal dimension, but it is easy to generalize this to arbitrary spacetime dimension and to three different conformal dimensions.}:
\bea
\tilde{g}^{(3)}_R(1,2;3)&=&{2^{5d/2+9-2\sum_i\delta_i}\pi^{2d+4}\over\Gamma(\delta_1)\Gamma(\delta_2)\Gamma(\delta_3)\Gamma(\sum_i\delta_i-{d\over2}-1)} \delta(\sum_i\omega_i)\delta(\sum_ip_{i,y})\delta^d(\sum_i\vec{p}_i)\int_0^\infty dzz^{d\over2} \nn\\&\times&\left[-(\omega_1-i\epsilon)^2+p_{1,y}^2+p_1^2\right]^{\nu_1/2}\mathrm{K}_{\nu_1}\left(z\sqrt{-(\omega_1-i\epsilon)^2+p_{1,y}^2+p_1^2}\right)\nn\\&\times& \left[-(\omega_2-i\epsilon)^2+p_{2,y}^2+p_2^2\right]^{\nu_2/2}\mathrm{K}_{\nu_2}\left(z\sqrt{-(\omega_2-i\epsilon)^2+p_{2,y}^2+p_2^2}\right)\nn\\&\times& \left[-(\omega_3+i\epsilon)^2+p_{3,y}^2+p_3^2\right]^{\nu_3/2}\mathrm{K}_{\nu_3}\left(z\sqrt{-(\omega_3+i\epsilon)^2+p_{3,y}^2+p_3^2}\right).
\eea
The notation here is a straightforward generalization of notation used in sections \ref{seczerotemptwopt} and \ref{secthreept}. The $\omega_i$, $p_{i,y}$ and $\vec{p}_i$ are relativistic energies and momenta. The spacetime is $d+2$ dimensional, where one of the spatial coordinates, $y$, is singled out. The third operator is singled out in $\tilde{g}^{(3)}_R(1,2;3)$ because, in position space, this correlator corresponds to the case where the operator $O_3$ has the largest time.  The $\delta_i$ and $\nu_i$ are as defined in (\ref{defofdelta}) and (\ref{defofnu}). Applying the map (\ref{momentummapinv}) to each of the $\omega_i$, $p_{i,y}$ pairs gives
\bea
\widetilde{G}^{(3)}_R(1,2;3)&=&{2^{5d/2+8-2\sum_i\delta_i}\pi^{2d+4}\over\Gamma(\delta_1)\Gamma(\delta_2)\Gamma(\delta_3)\Gamma(\sum_i\delta_i-{d\over2}-1)} \delta(\sum_i\Omega_i)\delta(\sum_iM_i)\delta^d(\sum_i\vec{p}_i)\int_0^\infty dzz^{d\over2} \nn\\&\times&\left[-4M_1\Omega_1+p_1^2+i\epsilon(\Omega_1+M_1)\right]^{\nu_1/2}\mathrm{K}_{\nu_1}\left(z\sqrt{-4M_1\Omega_1+p_1^2+i\epsilon(\Omega_1+M_1)}\right)\nn\\&\times& \left[-4M_2\Omega_2+p_2^2+i\epsilon(\Omega_2+M_2)\right]^{\nu_2/2}\mathrm{K}_{\nu_2}\left(z\sqrt{-4M_2\Omega_2+p_2^2+i\epsilon(\Omega_2+M_2)}\right)\nn\\&\times& \left[-4M_3\Omega_3+p_3^2-i\epsilon(\Omega_3+M_3)\right]^{\nu_3/2}\mathrm{K}_{\nu_3}\left(z\sqrt{-4M_3\Omega_3+p_3^2-i\epsilon(\Omega_3+M_3)}\right).\nn\\&&\label{retardedthreeptmomentum}
\eea
We will confirm in the next subsection that, in position space, $G_R^{(3)}(1,2;3)$ is such that $O_3$ has the largest time. The advanced non-relativistic three-point function is simply the complex conjugate of the retarded function:
\be
\widetilde G^{(3)}_A=(\widetilde G^{(3)}_R)^*.
\ee
These results will be used in the next subsection to compute retarded and advanced three-point correlators in position space.

As in the case of the time-ordered three-point function, we can extract a bulk-to-boundary propagator:
\be
\widetilde{\cal G}_R\sim z^{{d\over2}+1} \left[ -4M\Omega+p^2-i\epsilon(\Omega+M)\right]^{\nu/2}\mathrm{K}_{\nu}\left(z\sqrt{-4M\Omega+p^2-i\epsilon(\Omega+M)}\right).\label{retardedprop}
\ee
In inserting the factor $z^{{d\over2}+1}$, we have again anticipated the AdS/CFT result. This is a retarded non-relativistic ``bulk-to-boundary propagator". Notice that the three-point function $\widetilde G^{(3)}_R$ contains two different types of propagators which are distinguished by the sign in front of the $i\epsilon$. We have identified the retarded bulk-to-boundary propagator by selecting the propagator which is analytic in the upper half $\Omega$-plane. The other type of propagator appearing in $\widetilde G^{(3)}_R$ above is an advanced propagator:
\be
\widetilde{\cal G}_A\sim z^{{d\over2}+1} \left[ -4M\Omega+p^2+i\epsilon(\Omega+M)\right]^{\nu/2}\mathrm{K}_{\nu}\left(z\sqrt{-4M\Omega+p^2+i\epsilon(\Omega+M)}\right). \label{advancedprop}
\ee
The fact that the retarded three-point function can be expressed as an integral over two advanced propagators and one retarded propagator was pointed out in the relativistic case in \cite{BVWA}. Here we see that this same structure carries over to the non-relativistic theories we are considering. In section \ref{seczerotempadscft}, we will verify directly from non-relativistic AdS/CFT that it is appropriate to identify $\widetilde{\cal G}_R$ and $\widetilde{\cal G}_A$ as retarded and advanced bulk-to-boundary propagators by performing an analytic continuation from a Euclidean propagator.

Note that the real-time propagators $\widetilde{\cal G}_F$, $\widetilde{\cal G}_R$, and $\widetilde{\cal G}_A$ satisfy the relation
\be
\widetilde{\cal G}_F=\theta(\Omega+M)\widetilde{\cal G}_R+\theta(-\Omega-M)\widetilde{\cal G}_A.\label{GFGRGAproprelation}
\ee
This relation is not surprising when we recall that the momentum-space two-point correlators satisfy a similar relation (\ref{GFGRGArelationii}). Eqn.  (\ref{GFGRGAproprelation}) can be rewritten in a form analogous to (\ref{GFGRGArelation}),
\be
\widetilde{\cal G}_F=\theta(M)\widetilde{\cal G}_R+\theta(-M)\widetilde{\cal G}_A,\label{GFGRGAproprelationii}
\ee
as can be verified easily by rewriting the propagators as
\bea
\widetilde{\cal G}_F&\sim& e^{-i\pi\theta(-M)\nu/2}|M|^{\nu/2}\left(-2\Omega+{p^2\over2M}-i\epsilon M\right)^{\nu/2}\nn\\&\times&\mathrm{K}_\nu\left(z\sqrt{2|M|}e^{-i\pi\theta(-M)/2}\sqrt{-2\Omega+{p^2\over2M}-i\epsilon M}\right),\label{feynmanpropii}
\eea
\bea
\widetilde{\cal G}_R=\widetilde{\cal G}_A^*&\sim& e^{i\pi\theta(-M)\nu/2}|M|^{\nu/2}\left(-2\Omega+{p^2\over2M}-i\epsilon \right)^{\nu/2}\nn\\&\times&\mathrm{K}_\nu\left(z\sqrt{2|M|}e^{i\pi\theta(-M)/2}\sqrt{-2\Omega+{p^2\over2M}-i\epsilon }\right).\label{retardedadvancedprop}
\eea

\subsection{Retarded and advanced three-point functions in position space}\label{secretardedthreeptpos}
We will obtain the retarded and advanced three-point functions in position space by taking the Fourier transform of the momentum-space correlators computed in the previous subsection. As we saw, these correlators can be derived from relativistic three-point functions by redefining momentum space variables. The relativistic three-point function $\tilde g_R^{(3)}(1,2;3)$ which leads to the above expression for $\widetilde{G}_R^{(3)}(1,2;3)$ is the Fourier transform of the retarded three-point function for which the operator $O_3$ has the largest time. It is natural to expect that the same will be true for the non-relativistic three-point function $\widetilde{G}_R^{(3)}(1,2;3)$, and this is indeed the case. For simplicity, we restrict ourselves to the case $M_1>0$, $M_2>0$, which in turn implies that $M_3<0$. The details of the calculations are given in appendix \ref{appendixD}. For the retarded correlator, we find
\bea
\theta(M_1)\theta(M_2)G_R^{(3)}(1,2;3)&=&-{(2\pi)^3i^{\sum_i\delta_i}\delta(\sum_iM_i)\over\Gamma(\delta_1)\Gamma(\delta_2)\Gamma(\delta_3)}|x_{13}^+|^{-\delta_2} |x_{23}^+|^{-\delta_1} |x_{12}^+|^{-\delta_3} e^{iM_1 (x_{13}^2+i\epsilon)/x_{13}^+}\nn\\&\times&e^{iM_2(x_{23}^2+i\epsilon)/x_{23}^+}\theta(M_1)\theta(M_2)\theta(x_{31}^+)\theta(x_{32}^+)\bigg\{\theta(x_{12}^+) M_1^{\delta_2-1}M_2^{\delta_1+\delta_3-1}\nn\\&\times& B(\delta_3,\delta_1)\Phi_1(\delta_3,1-\delta_2,\delta_1+\delta_3,-{M_2\over M_1},-iv_{12}M_2)+\theta(x_{21}^+)\nn\\&\times&M_1^{\delta_2+\delta_3-1}M_2^{\delta_1-1}B(\delta_3,\delta_2)\Phi_1(\delta_3,1-\delta_1,\delta_2+\delta_3,-{M_1\over M_2},iv_{12}M_1)\bigg\},\nn\\&&\label{retardedthreeptposition}
\eea
with
\be
v_{12}\equiv{x_{12}^2\over x_{12}^+}+{x_{23}^2\over x_{23}^+}-{x_{13}^2\over x_{13}^+}.\label{vonetwodefiii}
\ee
The function $B(x,y)$ is the Euler beta function, while $\Phi_1$ is a hypergeometric function. Explicit definitions for these functions are given in appendix \ref{appendixA}. Notice that the above expression for $\theta(M_1)\theta(M_2)G_R^{(3)}$ is non-vanishing only when $x_3^+>x_2^+$ and $x_3^+>x_1^+$, which is consistent with our expectation that $G_R^{(3)}(1,2;3)$ is a retarded correlator for which $x_3^+/2$ is the largest time. The advanced correlator, with $x_3^+/2$ now being the smallest time, is found to be
\bea
\theta(M_1)\theta(M_2)G_A^{(3)}(1,2;3)&=&-{(2\pi)^3(-i)^{\sum_i\delta_i}\delta(\sum_iM_i)\over\Gamma(\delta_1)\Gamma(\delta_2)\Gamma(\delta_3)}|x_{13}^+|^{-\delta_2} |x_{23}^+|^{-\delta_1} |x_{12}^+|^{-\delta_3} e^{iM_1 (x_{13}^2+i\epsilon)/x_{13}^+}\nn\\&\times&e^{iM_2(x_{23}^2+i\epsilon)/x_{23}^+}\theta(M_1)\theta(M_2)\theta(x_{13}^+)\theta(x_{23}^+)\bigg\{\theta(x_{12}^+) M_1^{\delta_2+\delta_3-1}M_2^{\delta_1-1}\nn\\&\times& B(\delta_3,\delta_2)\Phi_1(\delta_3,1-\delta_1,\delta_2+\delta_3,-{M_1\over M_2},iv_{12}M_1)+\theta(x_{21}^+)\nn\\&\times&M_1^{\delta_2-1}M_2^{\delta_1+\delta_3-1}B(\delta_3,\delta_1)\Phi_1(\delta_3,1-\delta_2,\delta_1+\delta_3,-{M_2\over M_1},-iv_{12}M_2)\bigg\}.\nn\\&&\label{advancedthreeptposition}
\eea

In position space, retarded and advanced three-point functions can be expressed in terms of the Wightman (non-time-ordered) functions \cite{ls}. For instance, the retarded three-point function which corresponds to choosing $x^+_3$ to be the largest time is given by
\bea
G^{(3)}_R(x^+_1,x^+_2;x^+_3)&=&\theta(x^+_{31})\theta(x^+_{12})\left(G_{312}-G_{132}+G_{213}-G_{231}\right)\nn\\&+& \theta(x^+_{32})\theta(x^+_{21})\left(G_{321}-G_{231}+G_{123}-G_{132}\right).\label{retardedwightmanthreeptrelation}
\eea
We can use this formula to compute the retarded three-point correlator without making assumptions about the signs of the $M_i$. In the case $M_1>0$, $M_2>0$, it is easy to check using our formula for the Wightman function (\ref{wightmanthreeptposition}) that only $G_{123}$ and $G_{213}$ are non-vanishing, and that (\ref{retardedwightmanthreeptrelation}) agrees precisely with (\ref{retardedthreeptposition}).

We can find an alternate expression for the retarded three-point correlator in position space by taking the Fourier transform of the momentum space result (\ref{retardedthreeptmomentum}):
\bea
G_R^{(3)}(x^+_i,\vec{x}_i;M_i)&=&-{2^{4-{1\over2}\sum_i\nu_i}\pi^{6-{d\over2}}\Gamma(\Delta_1)\Gamma(\Delta_2)\Gamma(\Delta_3)\over\Gamma(\delta_1)\Gamma(\delta_2)\Gamma(\delta_3)\Gamma(\sum_i \delta_i-{d\over2}-1)}\delta(\sum_iM_i) \int_0^\infty dz \int dx^+d^dx z^{-d-3}\nn\\&\times&{\cal G}_R(x-x_1,z;M_1,\Delta_1){\cal G}_R(x-x_2,z;M_2,\Delta_2){\cal G}_R(x_3-x,z;M_3,\Delta_3),\nn\\&&
\eea
where
\be
{\cal G}_R(x,z;M,\Delta)={(-i)^{\Delta+1}\over2\pi\Gamma(\Delta)}|M|^{\Delta-1}z^\Delta\theta(x^+) |x^+|^{-\Delta}e^{iM(x^2+z^2)/x^+}
\ee
is the Fourier transform of the Schr\"{o}dinger retarded bulk-to-boundary propagator (\ref{retardedprop}). This result assumes that $M_1>0$, $M_2>0$, $M_3<0$.

\section{Zero-temperature bulk-to-boundary propagators in momentum space}\label{seczerotempadscft}
We will now confirm that our proposals for the real-time bulk-to-boundary propagators, (\ref{feynmanprop}), (\ref{retardedprop}), and (\ref{advancedprop}), are consistent with what are obtained from non-relativistic AdS/CFT. Since the results of the previous sections were derived without making recourse to AdS/CFT, it is reassuring that these results are reproduced when applying holographic duality prescriptions extended to non-relativistic theories \cite{Son,BM}. In AdS/CFT, we may construct retarded propagators by first computing Euclidean propagators and then performing an analytic continuation \cite{BVWA}. To arrive at a Schr\"{o}dinger-invariant theory, we must also perform (in the case of momentum space propagators or correlators) a change of momentum variables, as we saw in earlier sections. Once we have the retarded propagator, we may obtain the remaining real-time propagators from this by exploiting certain relations of which (\ref{GFGRGAproprelation}) is an example.

The zero-temperature Schr\"{o}dinger metric in Euclidean space Poincar\'{e}-like coordinates has the form \cite{HRR,ABM,MMT}
\be
ds^2_E=r^2(1+\beta^2r^2)d\tau^2-2i\beta^2r^4 d\tau dy +r^2(1-\beta^2r^2)dy^2+r^2dx^2+{dr^2\over r^2}.\label{HRRmetric}
\ee
We are following the notation of Herzog, Rangamani, and Ross \cite{HRR}. The radial coordinate $r$ ranges from zero to infinity, and the spacetime has a boundary at $r=\infty$. The Euclidean time $\tau$ and the Minkowski time $t$ are related by $\tau=-it$. We lump together $d$ of the spatial coordinates, $x_1,...,x_d$, into a $d$-dimensional vector $\vec{x}$ with $\vec{x}\cdot\vec{x}=x^2$; as in previous sections, we reserve vector notation for this $d$-dimensional subspace. Although the metric (\ref{HRRmetric}) is only a solution to (a truncated version of) type IIB supergravity when $d=2$, we will keep $d$ general throughout this section to facilitate comparison with the results of previous sections.

It is important to stress that $t$ should not be identified with physical time in the Schr\"{o}dinger field theory. The physical time\footnote{This coordinate was denoted $u$ in \cite{HRR}.} $x^+/2$ is obtained by switching to light cone coordinates:
\be
x^\pm=y\pm t.\label{light cone}
\ee
The time $t$ does admit the interpretation as the time coordinate of a parent theory that lives in one higher dimension relative to the Schr\"{o}dinger theory we wish to study. As we described earlier, the parent theory is a non-local dipole theory \cite{Alishahiha:2003ru,Bergman:2000cw,Dasgupta:2000ry}, and the Schr\"{o}dinger theory is obtained from this by applying DLCQ to the $x^-$ direction. Our approach will be to compute quantities in the parent theory and then perform a transformation (the momentum-space counterpart to (\ref{light cone})) to obtain quantities in the Schr\"{o}dinger theory.

We begin by computing the retarded propagator of a minimally coupled massive scalar $\phi$ in the parent theory. This can be obtained by first finding---in Euclidean signature---the solution $\phi_E$ to the scalar wave equation which is well behaved in the interior of the spacetime. After Fourier-transforming along all directions except $r$,
\be
F_E(\omega_E,p_y,\vec{p},r)=\int d\tau dy d^dx e^{-i\omega_E\tau-ip_yy-i\vec{p}\cdot\vec{x}}\phi_E(\tau,y,\vec{x},r),\label{FT}
\ee
the wave equation has the form:
\be
F_E''+{d+3\over r}F_E'-{1\over r^4}\left[\omega_E^2+p_y^2+p^2+\beta^2(p_y+i\omega_E)^2r^2+m^2r^2\right]F_E=0.
\ee
The prime denotes differentiation with respect to $r$, and $m$ is the mass of the scalar. The solution to this equation which is well behaved as $r\to0$ is
\be
F_E(\omega_E,p_y,\vec{p},r)={C\over r^{{d\over2}+1}}K_{\nu}(k/r),\label{euclidprop}
\ee
with
\be
k\equiv \sqrt{\omega_E^2+p_y^2+p^2},\label{euclidk}
\ee
and
\be
\nu\equiv \sqrt{{(d+2)^2\over4}+m^2+\beta^2(p_y+i\omega_E)^2}.\label{euclidnu}
\ee
The constant $C$ can be fixed by normalizing the solution at some large but finite value of $r$. If we now analytically continue $\omega_E\to-i(\omega+i\epsilon)$, then the resulting function will be analytic in the upper-half $\omega$ plane and will thus admit the interpretation as a retarded propagator in the parent theory.

We would like to re-interpret this result in terms of the Schr\"{o}dinger theory. We do this by first inverting the coordinate transformation (\ref{light cone}):
\be
\tau={-i\over2}(x^+-x^-),\qquad y={1\over2}(x^++x^-).
\ee
Applying this coordinate transformation to (\ref{FT}) allows us to identify the frequency $\Omega$ associated with the time coordinate $x^+$ and the momentum $M$ associated with the coordinate $x^-$:
\be
\Omega={1\over2}(\omega-p_y),\qquad M={1\over2}(\omega+p_y).\label{light conemomenta}
\ee
The inverse of this map is
\be
\omega=\Omega+M,\qquad p_y=M-\Omega.\label{light conemomentainv}
\ee
These mappings are of course identical to (\ref{momentummap}) and (\ref{momentummapinv}). The retarded propagator parameters can then be rewritten as
\be
k=\sqrt{-4M\Omega +p^2-i\epsilon\left(\Omega+M\right)},
\ee
\be
\nu=\sqrt{{(d+2)^2\over4}+m^2+4\beta^2M^2}.\label{numasses}
\ee
Setting $r=1/z$, we then have
\be
F_E\to \widetilde{\cal G}_R= C z^{{d\over2}+1} \mathrm{K}_{\nu}\left(z\sqrt{-4M\Omega+p^2-i\epsilon(\Omega+M)}\right),\label{retardedpropadscft}
\ee
which is consistent with our earlier finding for the retarded bulk-to-boundary propagator, eqn. (\ref{retardedprop}), if we identify the conformal dimension $\Delta$ according to
\be
\Delta={d\over2}+1+\nu={d\over2}+1+\sqrt{{(d+2)^2\over4}+m^2+4\beta^2M^2}.\label{conformaldim}
\ee

We note the similarity between the AdS scalar bulk-to-boundary propagator and the scalar propagator given in (\ref{euclidprop}), (\ref{euclidk}), and (\ref{euclidnu}), with the main difference being the extra term $\beta^2(p_y+i\omega_E)^2$ inside the square root in (\ref{euclidnu}). The AdS case corresponds, of course, to having $\beta=0$. This change in $\nu$ is the only effect of having a nonzero $\beta$, at least as far as the scalar bulk-to-boundary propagators are concerned.

We can apply this procedure to the other types of real-time bulk-to-boundary propagators identified in \cite{BVWA}. One finds for instance that the Feynman and advanced bulk-to-boundary propagators constructed in this way agree precisely with (\ref{feynmanprop}) and (\ref{advancedprop}), again writing the conformal dimension as in (\ref{conformaldim}). It was argued in \cite{BVWA} that one can consistently define Wightman bulk-to-boundary propagators by extending the applicability of standard flat space circling rules. This approach naturally commutes to the present context, with the Wightman propagators given by
\be
\widetilde{\cal G}^+ = \widetilde{\cal G}_F-\widetilde{\cal G}_A,\qquad \widetilde{\cal G}^-=\widetilde{\cal G}_F-\widetilde{\cal G}_R.
\ee
Combining these expressions with (\ref{GFGRGAproprelation}), it is straightforward to derive that the Wightman propagators can also be written as
\be
\widetilde{\cal G}^\pm = i\left[\mathrm{sgn}(\Omega+M)\pm1\right]\mathrm{Im}\widetilde{\cal G}_R.
\ee

It is easy to check that the various bulk-to-boundary propagators identified in this and the previous section reproduce the momentum-space two-point functions computed in section \ref{seczerotemptwopt}. This essentially follows from standard AdS/CFT two-point function computations. Since the momenta play a passive role in the AdS/CFT prescription which takes us from bulk-to-boundary propagator to two-point correlator, it makes no difference whether we apply the mapping (\ref{light conemomentainv}) before or after we execute this prescription. Moreover, the different $i\epsilon$ prescriptions which distinguish the various types of bulk-to-boundary propagator are also passive and carry over directly to the resulting two-point functions.

\section{Finite-temperature correlators}\label{secfiniteTadscft}

\subsection{Black hole metric and bulk-to-boundary propagators}\label{secfiniteTprops}
Finally, we turn to computing Schr\"{o}dinger correlation functions at finite temperature from non-relativistic AdS/CFT. The previous sections have established the various ingredients we will need to obtain these correlators, in particular the momentum-space mapping (\ref{light conemomentainv}) from the parent to the Schr\"{o}dinger theories and the identification of the different real-time propagators. We will begin by using these methods to construct the different real-time bulk-to-boundary propagators at finite temperature. We will first obtain the retarded propagator by computing a thermal Euclidean propagator and then analytically continuing this and using (\ref{light conemomentainv}) to obtain a thermal retarded propagator for the gravity dual of the Schr\"{o}dinger theory. The remaining real-time propagators can be obtained from this using various identities borrowed from the case of a relativistic CFT. With these propagators in hand, we will then proceed to compute various two-point correlators as well as the retarded and time-ordered three-point correlators with the help of circling rules as in \cite{BVWA}.

To study the theory at finite temperature, we consider the following (Euclidean) black hole geometry \cite{HRR,ABM,MMT}:
\bea
ds^2_E&=&r^2k(r)^{-2/3}\left[-\beta^2r^2f(r)(id\tau+dy)^2+f(r)d\tau^2+dy^2+k(r)dx^2\right]\nn\\&+&k(r)^{1/3}{dr^2\over r^2f(r)},
\eea
\be
f(r)\equiv 1-{r_+^4\over r^4},\quad k(r)\equiv 1+{\beta^2r_+^4\over r^2}.
\ee
$r_+$ is the location of the event horizon. As in the case of zero temperature, we are adopting most of the notational conventions of \cite{HRR}. In this section, we restrict attention to the case $d=2$ (i.e. $\vec{x}$ is two-dimensional). In order to simplify some of the expressions that will follow and to make contact with previous work, we will define a new radial coordinate $u$:
\be
u\equiv {r_+^2\over r^2}.
\ee
The metric then reads
\bea
ds^2_E&=&{r_+^2\over u}k(u)^{-2/3}\left[-\beta^2r_+^2{f(u)\over u}(id\tau+dy)^2+f(u)d\tau^2+dy^2+k(u)dx^2\right]\nn\\&+&k(u)^{1/3}{du^2\over4u^2f(u)},\label{bhmetric}
\eea
\be
f(u)=1-u^2,\qquad k(u)=1+\beta^2r_+^2u.
\ee
The temperature can be computed from the surface gravity, which in turn is computed from the Killing vector generator of the event horizon, $\xi$, according to the formula
\be
\kappa^2=-{1\over2}\left(\nabla^a\xi^b\right)\left(\nabla_a\xi_b\right).
\ee
The vector $\xi$ is normalized so that the component along the Schr\"{o}dinger time direction is 1:
\be
\xi\equiv 2{\partial\over\partial t}=2{\partial\over\partial x^+}-2{\partial\over\partial x^-}.\label{nullhorizongen}
\ee
With this definition, we find that the temperature of the Schr\"{o}dinger theory is given by twice that of the parent dipole theory,
\be
T_S={\kappa\over2\pi}={2r_+\over\pi}=2T_{p},
\ee
since $T_{p}$ should be measured with respect to the time $t$. From (\ref{nullhorizongen}), we see that we also have a chemical potential $\mu$ associated with the conserved charge $\partial\over\partial x^-$. With our set of conventions,
\be
\mu=-2.
\ee
As was noted in \cite{HRR,ABM}, these observations can also be discerned by considering a canonical ensemble for the parent theory with a density matrix given by
\be
\rho=e^{-H_{p}/ T_{p}}= e^{-[{1\over2}H_{S}-P_-]/ T_{p}}=e^{-[H_{S}-2P_-]/T_S}=e^{-[H_{S}+\mu P_-]/T_S},
\ee
where the Hamiltonian $H_{p}={\partial\over\partial t}$, $H_{S}=2{\partial\over\partial x^+}$, and $P_-={\partial\over\partial x^-}$. We see that the thermal  Schr\"{o}dinger theory is naturally described in terms of a grand canonical ensemble, where $P_-$ is interpreted as the particle number operator.

We will focus on the case of a minimally coupled {\it massless} scalar for simplicity\footnote{An example of such a scalar is the fluctuation in the metric component $g_{x_1}^{x_2}$, which has been shown to decouple from all other fluctuations \cite{HRR,ABM}. As we show in appendix \ref{appendixE}, this scalar fluctuation can be made massive by considering non-trivial harmonics on the internal 5d space, in which case the wave equation generalizes to $F_E''-{1+u^2\over u(1-u^2)}F_E'-{1\over u^2(1-u^2)^2}\bigg\{\gomega_E^2u+(\p_y^2+\p^2+{1\over4}\Lambda^2n_\chi^2)u(1-u^2)+[\Lambda^2(\p_y+i\gomega_E)^2+{1\over4}l(l+4)](1-u^2)\bigg\}F_E=0$, where $l$ and $n_\chi$ are quantum numbers defined in the appendix.}. In momentum space, the equation of motion has the form:
\bea
&&F_E''-{1+u^2\over u(1-u^2)}F_E'-{1\over u^2(1-u^2)^2}\bigg[\gomega_E^2u+(\p_y^2+\p^2)u(1-u^2)\nn\\&&+\Lambda^2(\p_y+i\gomega_E)^2(1-u^2)\bigg]F_E=0,
\eea
where we have defined
\be
\Lambda\equiv\beta r_+.
\ee
We have also defined the dimensionless quantities $\gomega_E$, $\p_y$, and $\vec{\p}$ according to
\be
\gomega_E\equiv {\omega_E\over 2r_+},\quad \p_y\equiv{p_y\over 2r_+},\quad \vec{\p}\equiv {\vec{p}\over2r_+}.
\ee
To solve the equation of motion, we use the following ansatz for the Euclidean propagator:
\be
F_E(\gomega_E,\p_y,\vec{\p},u)=u^\lambda(1-u)^{\gomega_E/2}(1+u)^{i\gomega_E/2} H(\gomega_E,\p_y,\vec{\p},u).
\ee
The resulting equation for $H$ is the Heun equation,
\be
H''+\left({\gamma\over u}+{\delta\over u-1}+{\varepsilon\over u-\hd}\right)H'+{\alpha\beta u-q\over u(u-1)(u-\hd)}H=0,\label{heuneqn}
\ee
so long as we take\footnote{We have chosen the root which reproduces the zero-temperature limit, in which $\Lambda\to0$ and $\lambda\to0$.}
\be
\lambda=1-\sqrt{1+\Lambda^2(\p_y+i\gomega_E)^2}.
\ee
The Heun parameters are
\bea
&&\gamma=1-2\sqrt{1+\Lambda^2(\p_y+i\gomega_E)^2},\quad \delta=1+\gomega_E,\quad \varepsilon=1+i\gomega_E,\nn\\ &&\alpha=\beta={1+i\over2}\gomega_E+1-\sqrt{1+\Lambda^2(\p_y+i\gomega_E)^2},\quad \hd=-1,\nn\\ && q={1-i\over2}\gomega_E\left[2\sqrt{1+\Lambda^2(\p_y+i\gomega_E)^2}-1\right]-\gomega_E^2-\p_y^2-\p^2.
\eea
The Heun's function $H$ must be chosen to be the solution which is well behaved near the origin $u=1$ of the Euclidean spacetime, which for positive frequencies is
\footnote{See ref. \cite{192} for a comprehensive discussion of the Heun equation and its solutions. A brief review of aspects essential for the present context is also given in appendix E of \cite{BVWA}.}:
\be
H(\gomega_E,\p_y,\vec{\p},u)=Hl(1-\hd,\alpha\beta-q;\alpha,\beta,\delta,\gamma;1-u).\label{heunsoln}
\ee
Analytically continuing $\gomega_E\to-i(\gomega+i\epsilon)$, using (\ref{light conemomentainv}), and defining $\gOmega\equiv\Omega/(2r_+)$ and $\M\equiv M/(2r_+)$, we find that the retarded propagator is\footnote{It is amusing to note that if we keep the $i\epsilon$ in the exponent as well so that $\widetilde{\cal G}_R\sim (1-u)^{-{i\over2}(\gOmega+\M+i\epsilon)}$, then the retarded propagator vanishes at the horizon $u=1$ as opposed to the usual boundary condition that it behave like an in-coming wave.}
\be
\widetilde{\cal G}_R(\gOmega,\vec{\p},u;\M)\sim u^\lambda(1-u)^{-{i\over2}(\gOmega+\M)}(1+u)^{{1\over2}(\gOmega+\M)}H(\gOmega,\vec{\p},u),
\ee
with
\bea
&&\lambda=1-\sqrt{1+4\M^2\Lambda^2},\quad \gamma=1-2\sqrt{1+4\M^2\Lambda^2},\quad \delta=1-i\left(\gOmega+\M+i\epsilon\right),\nn\\ && \varepsilon=1+\gOmega+\M+i\epsilon,\nn\\ && \alpha=\beta= {1-i\over2}\left(\gOmega+\M+i\epsilon\right)+1-\sqrt{1+4\M^2\Lambda^2},\nn\\ && q=-{1+i\over2}\left(\gOmega+\M+i\epsilon\right)\left(2\sqrt{1+4\M^2\Lambda^2}-1\right)+ 4\M\gOmega -\p^2+i\epsilon\left(\gOmega+\M\right).\nn\\&&
\eea
Since it is not possible to normalize $\widetilde{\cal G}_R$ at $u=0$, we introduce a cutoff $u_B\ll1$, and normalize the propagator at $u=u_B$:
\be
\widetilde{\cal G}_R(\gOmega,\vec{\p},u;\M)={u^\lambda(1-u)^{-{i\over2}(\gOmega+\M)}(1+u)^{{1\over2}(\gOmega+\M)}\over u_B^\lambda(1-u_B)^{-{i\over2}(\gOmega+\M)}(1+u_B)^{{1\over2}(\gOmega+\M)}}{H(\gOmega,\vec{\p},u)\over H(\gOmega,\vec{\p},u_B)}.\label{thermalretardedprop}
\ee

As in \cite{BVWA}, we define the thermal Feynman bulk-to-boundary propagator in the black hole background to be given by
\be
\tilde{\mathfrak g}_F(\gomega,\p_y,\vec{\p},u)=\mathrm{Re}\tilde{\mathfrak g}_R(\gomega,\p_y,\vec{\p},u)+i\coth\left(\pi\gomega\right)\mathrm{Im}\tilde{\mathfrak g}_R(\gomega,\p_y,\vec{\p},u).
\ee
Although \cite{BVWA} focused primarily on the gravity dual of thermal ${\cal N}{=}4$ super Yang-Mills theory, the arguments (which were based on extending field theory finite-temperature circling rules to bulk-to-boundary propagators) were quite general and should apply to a wide class of theories including the dipole theory conjectured to be dual to the black hole spacetime (\ref{bhmetric}). (Note that thermal ${\cal N}{=}4$ SYM corresponds to the special case $\Lambda=0$.) Applying (\ref{light conemomentainv}), we obtain a similar relation for the Schr\"{o}dinger propagators as well:
\be
\widetilde{\cal G}_F(\gOmega,\vec{\p},u;\M)=\mathrm{Re}\widetilde{\cal G}_R(\gOmega,\vec{\p},u;\M)+i\coth\left(\pi(\gOmega+\M)\right)\mathrm{Im}\widetilde{\cal G}_R(\gOmega,\vec{\p},u;\M).\label{finiteTGFGRrelation}
\ee
This formula allows us to compute the Feynman bulk-to-boundary propagator directly from the retarded propagator we have already calculated. Again borrowing results from the relativistic case, we may compute Wightman propagators using the same relations we have at zero temperature \cite{BVWA}:
\be
\widetilde{\cal G}^+ = \widetilde{\cal G}_F-\widetilde{\cal G}_A,\qquad \widetilde{\cal G}^-=\widetilde{\cal G}_F-\widetilde{\cal G}_R.\label{finiteTwightmanrelations}
\ee
In conjuction with (\ref{finiteTGFGRrelation}), these expressions imply that
\be
\widetilde{\cal G}^\pm={\pm 2i\over 1-e^{\mp2\pi(\gOmega+\M)}}\mathrm{Im}\widetilde{\cal G}_R.\label{finiteTwightmanprop}
\ee

\subsection{Real-time thermal two-point correlators}\label{secfiniteTtwopt}
The real-time bulk-to-boundary propagators obtained in the previous subsection can be used to compute real-time two-point correlators at finite temperature. We start with the retarded two-point function, which arises from a boundary term in the quadratic part of the effective 5d action\footnote{We may take the effective 5d action to be any of a number of consistent truncations of IIB supergravity that have been found in the literature \cite{HRR,ABM,MMT,Liu:2010sa,Gauntlett:2010vu,Skenderis:2010vz,Cassani:2010uw}. Since we are considering a generic minimally coupled scalar, our results are independent of this choice.} of the form $\int d\Omega d^2p\sqrt{-g}g^{uu} \phi(-\Omega,-\vec{p})\partial_u\phi(\Omega,\vec{p})|_{u=u_B}$. Setting the scalar field $\phi(\Omega,\vec{p},u)=\widetilde{\cal G}_R(\Omega,\vec{p},u)\phi_B(\Omega,\vec{p})$ and varying with respect to the boundary value $\phi_B$, we find
\be
\widetilde{G}_R(\Omega,\vec{p};M)=-2\bar N \sqrt{-g}g^{uu}\widetilde{\cal G}_R(\Omega,\vec{p},u;M)\partial_u\widetilde{\cal G}_R(\Omega,\vec{p},u;M)\big|_{u=u_B}.\label{twoptformula}
\ee
$\bar N$ is a supergravity normalization factor. In order to evaluate this expression, we must expand the Heun's function $H$ near the boundary at $u=u_B$. This is facilitated by first expressing the Heun's function (\ref{heunsoln}) as a linear combination of independent solutions defined in the vicinity of $u=0$:
\be
H(u)=H_1(u)+A H_2(u),\label{heunsolnii}
\ee
where
\be
H_1(u)=Hl(\hd,q,\alpha,\beta,\gamma,\delta;u),
\ee
and
\be
H_2(u)=u^{1-\gamma}Hl(\hd,q-(\gamma-1)(\delta \hd+\varepsilon),\beta-\gamma+1,\alpha-\gamma+1,2-\gamma,\delta;u).
\ee
The coefficient $A$ is determined by matching (\ref{heunsoln}) and (\ref{heunsolnii}) at some value of $u$ between 0 and 1; in practice, this must be done numerically. Since our formula for the two-point function, eqn. (\ref{twoptformula}), only depends on the behavior of $\widetilde{\cal G}_R$ and its first derivative near $u=u_B\ll1$, it suffices to expand $H_1$ and $H_2$ about $u=0$:
\be
H_1(u)=1+c_1u+c_2u^2+...,\qquad H_2=u^2+...,
\ee
with\footnote{Note that if $\gamma$ is a negative integer, then $H_1$ can no longer be expanded in a simple power series---logarithmic factors will also appear. See appendix E of \cite{BVWA} for a more complete discussion of this subtlety.}
\be
c_1=-{q\over\gamma},\qquad c_2={q(\varepsilon-\delta+q)+\alpha\beta\gamma\over2\gamma(\gamma+1)},
\ee
where we have set the Heun parameter $\hd=-1$. After throwing away terms which diverge in the $u_B\to0$ limit, we find
\be
\widetilde{G}_R(\gOmega,\vec{\p};\M)=4r_+^4\bar N\left[\lambda-2(A+c_2)+c_1^2+{1-i\over2}(\gOmega+\M)\right].\label{thermalretardedtwopt}
\ee
We can obtain an analytic expression for $A$ in the limit of small $\gOmega$, $\M$, and $\vec{\p}$ by plugging the ansatz $H(u)=1+q h(u)$ into (\ref{heuneqn}) and expanding the Heun parameters. The solution which is regular at $u=1$, namely $h(u)=\log(1+u)$, gives a solution $H(u)$ which is valid for all $u$ so long as we only keep the leading order terms in $q$. Expanding this solution to second order in $u$ about $u=0$ then gives
\be
A+c_2={1+i\over 4}(\gOmega+\M)+{1\over2}\p^2+O(\gOmega^2,\gOmega \M,\M^2,\gOmega \p^2,\M \p^2,\p^4).
\ee
Plugging this into (\ref{thermalretardedtwopt}) yields the thermal retarded two-point correlator in the limit of small frequency and mass. This quantity was also obtained in \cite{ABM} for the case $\vec{p}=0$.

To compute the Feynman two-point function, we can
perform a similar analysis employing the various thermal
bulk-to-boundary propagators along with finite temperature
circling rules. We will now give a brief review of circling rules as they apply to Witten diagrams. Each vertex of the diagram can be either circled or uncircled, and a general real-time correlator will be given by a sum of diagrams having particular vertices circled. The circlings determine the type of real-time correlator (e.g. time-ordered, retarded, etc.): if $\widetilde{\cal G}_i$ extends between two uncircled vertices, then $\widetilde{\cal G}_i=\widetilde{\cal G}_F$, the Feynman propagator. In the case of two circled vertices, we instead have $\widetilde{\cal G}_i=\widetilde{\bar{\cal G}}_F=-\widetilde{\cal G}_F^*$. Since the momenta $p_i$ are defined to all be out-going, $\widetilde{\cal G}_i=\widetilde{\cal G}^-$ when the bulk vertex is circled and the boundary vertex is uncircled, and $\widetilde{\cal G}_i=\widetilde{\cal G}^+$ when the reverse is true. If the number of circled vertices in a given diagram is odd, then the contribution from that diagram also receives an overall minus sign. In particular, for the time-ordered two-point function, the vertices at the boundary are uncircled, and the derivative-coupling bulk vertex can be either circled or un-circled. Summing over both contributions yields
\bea
\widetilde{G}_F(\Omega,\vec{p};M)&=&-2\bar N \sqrt{-g}g^{uu}\bigg(\widetilde{\cal G}_F(\Omega,\vec{p},u;M)\partial_u\widetilde{\cal G}_F(\Omega,\vec{p},u;M)\nn\\
&-&\widetilde{\cal G}^-(\Omega,\vec{p},u;M)\partial_u\widetilde{\cal G}^+(\Omega,\vec{p},u;M)\bigg)\Big|_{u=u_B},
\eea
which, upon substituting (\ref{finiteTGFGRrelation})  and (\ref{finiteTwightmanprop}),  leads to
\be
\widetilde{G}_F(\gOmega,\vec{\p};\M)=\mathrm{Re}\widetilde{G}_R(\gOmega,\vec{\p};\M)+i\coth\left(\pi(\gOmega+\M)\right)\mathrm{Im}\widetilde{ G}_R(\gOmega,\vec{\p};\M).
\ee
Recalling that $\gOmega={\Omega\over2r_+}$ and $\M={M\over2r_+}$, where $\Omega$ and $M$ are the physical (dimensionful) frequency and mass, we see that this is precisely the standard Kallen-Lehmann relation at finite temperature with energy $2\Omega$, particle number $M$, and chemical potential $\mu=-2$. It is straightforward to check that in the zero-temperature limit where $r_+\to0$, this reduces to the relation we found in section \ref{secretardedtwoptmom}, eqn. (\ref{GFGRGArelationii}).

A similar application of circling rules in the case of the Wightman functions leads to
\be
\widetilde{G}^\pm = {\pm 2i\over 1-e^{\mp2\pi(\gOmega+\M)}}\mathrm{Im}\widetilde{G}_R.
\ee

\subsection{Thermal retarded and advanced three-point correlators}\label{secfiniteTretardedthreept}
The retarded three-point scalar correlator (which is  equal to the sum of all diagrams with vertices either circled or uncircled with the exception of the vertex with the largest time which remains uncircled) has the same universal expression in the black hole background which is dual to the dipole theory as the one obtained in \cite{BVWA}:
\be
\tilde g_R^{(3)}(p_1,p_2;p_3)\sim\delta^4(p_1+p_2+p_3)\int_{u_B}^1 du\sqrt{-g}\tilde{\mathfrak g}_R^*(p_1,u)\tilde{\mathfrak g}_R^*(p_2,u)\tilde{\mathfrak g}_R(p_3,u).\label{relativisticthreept}
\ee
The position-space version of this correlator is such that the operator $O_3$ has the largest time. Note that the four-momenta $p_i$ are defined to be all out-going, i.e. they point toward the boundary. This result can be obtained in a number of ways, including the application of circling rules to Witten diagrams (as advertised) or by analytically continuing the Euclidean three-point function.

At this point, we should make a few comments regarding the integration region in (\ref{relativisticthreept}). In the case of thermal ${\cal N}{=}4$ SYM, it has been proposed that one must consider the maximally extended AdS-Schwarzschild black hole spacetime in order to reproduce Schwinger-Keldysh correlators from AdS/CFT \cite{HerzogSon}. That is, in addition to the quadrant which is integrated over in (\ref{relativisticthreept}), there are three more quadrants that one should also take into consideration. However, as is argued in \cite{BVWA}, the use of circling rules renders unnecessary the introduction of additional quadrants. This result is useful since it avoids one of the more subtle issues which arise in real-time AdS/CFT calculations, namely the role of global properties of the bulk spacetime. We will keep exploiting this simplification. It was also pointed out in \cite{BVWA} that in the case of a retarded three-point correlator, neglecting extra quadrants is consistent with the expectation that one should need to integrate only over a maximally causal region of the spacetime that includes the original boundary at $u{=}u_B$. (In \cite{BVWA}, $u_B{=}0$.) This is tantamount to integrating over only the original quadrant---from the event horizon at $u{=}1$ to the boundary at $u{=}u_B$---since the other three are causally disconnected from it, as is readily apparent from the fact that the remaining quadrants lie beyond the event horizon of the black hole. ( For a different perspective reaching the same conclusion see \cite{vanRees:2009rw}.)

Having argued for the validity of the three-point function (\ref{relativisticthreept}) for the parent theory, it is then a simple matter to transform this to a Schr\"{o}dinger three-point function using (\ref{light conemomentainv}) for each of the three momenta:
\bea
\widetilde G_R^{(3)}(\Omega_i,\vec{p}_i;M_i)&\sim&\delta(\sum_iM_i)\delta(\sum_i\Omega_i)\delta(\sum_i\vec{p}_i)\nn\\&\times&\int_{u_B}^1 du\sqrt{-g}\widetilde{\cal G}_R^*(\Omega_1,\vec{p}_1,u;M_1)\widetilde{\cal G}_R^*(\Omega_2,\vec{p}_2,u;M_2)\widetilde{\cal G}_R(\Omega_3,\vec{p}_3,u;M_3).\nn\\&&\label{finiteTnonrelthreept}
\eea
The thermal retarded bulk-to-boundary propagator $\widetilde{\cal G}_R$ was given in (\ref{thermalretardedprop}). The thermal advanced three-point function is given by the complex conjugate of (\ref{finiteTnonrelthreept}).

\subsection{Thermal time-ordered three-point correlator}\label{secfiniteTfeynmanthreept}
In this section, we compute the Schr\"{o}dinger thermal time-ordered three-point correlator. As we have done in the case of the retarded three-point in the previous subsection, we could first compute the time-ordered three-point function in the parent theory and then transform this using (\ref{light conemomentainv}) to obtain the Schr\"{o}dinger time-ordered function. However, considering the various results we have obtained in earlier sections, it is clear that this will be equivalent to a direct computation using a non-relativistic version of Witten diagrams and circling rules. A generic three-point Witten diagram is of the form
\be
\delta(\sum_iM_i)\delta(\sum_i\Omega_i)\delta(\sum_i\vec{p}_i)\int_{u_B}^1 du\sqrt{-g}\widetilde{\cal G}_a(\Omega_1,\vec{p}_1,u;M_1)\widetilde{\cal G}_b(\Omega_2,\vec{p}_2,u;M_2)\widetilde{\cal G}_c(\Omega_3,\vec{p}_3,u;M_3),\label{genericdiagram}
\ee
where the types of bulk-to-boundary propagators $\widetilde{\cal G}_i$ appearing in (\ref{genericdiagram}) depend on which vertices are circled.
\begin{figure}
\begin{center}
\includegraphics[width=4.0in]{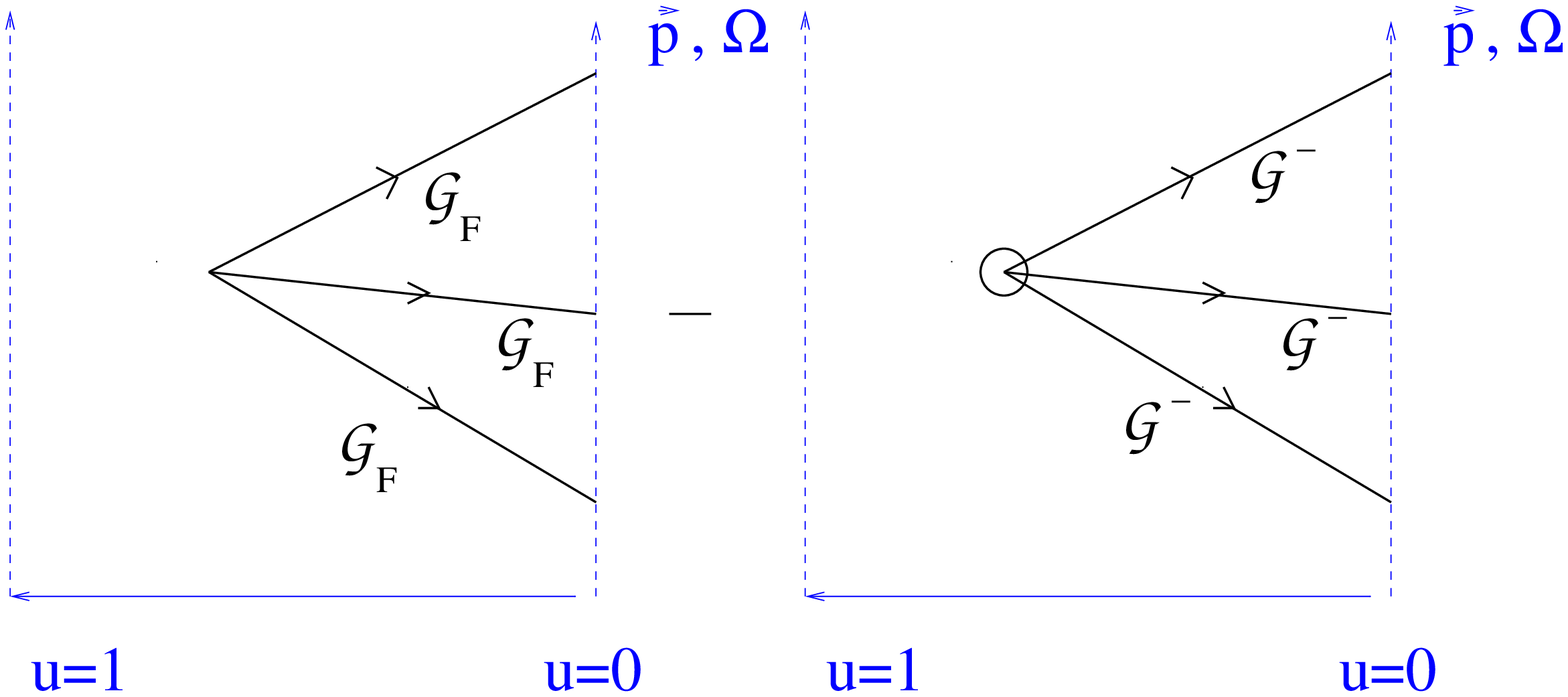} \\
Figure 1: Witten diagrams contributing to the time-ordered three-point correlator.
\end{center}
\end{figure}

In the case of a time-ordered function, we must add the contributions of all diagrams which result from considering all possible circlings of internal vertices, with the external vertices uncircled. For a three-point function tree-level diagram there is only one internal vertex, and the two contributing diagrams are shown in Fig. 1. A straightforward application of the circling rules reviewed above gives
\bea
\widetilde G_F^{(3)}(\Omega_i,\vec{p}_i)&\sim&\delta(\sum_iM_i)\delta(\sum_i\Omega_i)\delta(\sum_i\vec{p}_i)\bigg\{\nn\\&\times&\int_{u_B}^1 du\sqrt{-g}\widetilde{\cal G}_F(\Omega_1,\vec{p}_1,u;M_1)\widetilde{\cal G}_F(\Omega_2,\vec{p}_2,u;M_2)\widetilde{\cal G}_F(\Omega_3,\vec{p}_3,u;M_3)\nn\\&-&\int_{u_B}^1 du\sqrt{-g}\widetilde{\cal G}^-(\Omega_1,\vec{p}_1,u;M_1)\widetilde{\cal G}^-(\Omega_2,\vec{p}_2,u;M_2)\widetilde{\cal G}^-(\Omega_3,\vec{p}_3,u;M_3)\bigg\}.\nn\\&&\label{finiteTnonrelfeynmanthreept}
\eea
This result is consistent with (\ref{feynmanthreeptmomentum}) in the zero-temperature limit since in this limit, the second term vanishes. This can be seen from (\ref{finiteTwightmanprop}) by writing $\gOmega+\M={\Omega+M\over 2r_+}$ and sending $r_+\to0$ while keeping $\Omega$ and $M$ fixed. Momentum conservation ensures that $\Omega_i+M_i<0$ for at least one of the three sets of momenta, and the corresponding Wightman propagator $\widetilde{\cal G}^-(\Omega_i,\vec{p}_i,u;M_i)$ will therefore vanish.

\bigskip
\bigskip
\bigskip
\noindent {\Large\bf Acknowledgements}

\noindent We would like to thank Leopoldo Pando Zayas and Djordje Minic for helpful discussions. This work was supported in part by Jeffress research grant GF12334.
\bigskip

\begin{appendix}
\section{Zero-temperature time-ordered three-point functions in position space}\label{appendixA}
We begin with the relativistic conformal three-point correlator:
\bea
g^{(3)}_F&=&-\langle0|{\cal T}O_1(x_1)O_2(x_2)O_3(x_3)|0\rangle=-{1\over[x_{23}^+x_{23}^-+x_{23}^2+i\epsilon]^{\delta_1}}\nn\\&\times&{1\over[x_{13}^+x_{13}^-+x_{13}^2+i\epsilon]^{\delta_2}} {1\over[x_{12}^+x_{12}^-+x_{12}^2+i\epsilon]^{\delta_3}},
\eea
where $x_{ij}=x_i-x_j$. Fourier-transforming with respect to $x_i^-$ and introducing Schwinger parameters gives
\bea
G^{(3)}_F&=&-{(-i)^{\delta_1+\delta_2+\delta_3}\over\Gamma(\delta_1)\Gamma(\delta_2)\Gamma(\delta_3)}\int_0^\infty ds_1ds_2ds_3 s_1^{\delta_1-1}s_2^{\delta_2-1}s_3^{\delta_3-1}\int dx_1^-dx_2^-dx_3^- e^{-iM_1x_1^--iM_2x_2^--iM_3x_3^-}\nn\\&\times& e^{-s_1[\epsilon-ix_{23}^+x_{23}^--ix_{23}^2]} e^{-s_2[\epsilon-ix_{13}^+x_{13}^--ix_{13}^2]}e^{-s_3[\epsilon-ix_{12}^+x_{12}^--ix_{12}^2]}.
\eea
It is useful to change integration variables from $x_i^-$ to $x_{13}^-$, $x_{23}^-$, and $x_3^-$. The $x_3^-$ integration then gives a $\delta$-function of the form $\delta(M_1+M_2+M_3)$. The $x_{13}^-$ and $x_{23}^-$ integrals also produce $\delta$-functions, and we obtain
\bea
G^{(3)}_F&=&-{(2\pi)^3(-i)^{\delta_1+\delta_2+\delta_3}\delta(\sum_iM_i)\over\Gamma(\delta_1)\Gamma(\delta_2)\Gamma(\delta_3)}\int_0^\infty ds_1ds_2ds_3 s_1^{\delta_1-1}s_2^{\delta_2-1}s_3^{\delta_3-1}\nn\\&\times& e^{i(x_{23}^2+i\epsilon)s_1}e^{i(x_{13}^2+i\epsilon)s_2}e^{i(x_{12}^2+i\epsilon)s_3}\delta(-M_1+x_{12}^+s_3+x_{13}^+s_2)\delta(-M_2-x_{12}^+s_3+x_{23}^+s_1).\nn\\&&
\eea
Evaluating the $\delta$-functions, we find
\bea
G^{(3)}_F&=&-{(2\pi)^3(-i)^{\delta_1+\delta_2+\delta_3}\delta(\sum_iM_i)\over\Gamma(\delta_1)\Gamma(\delta_2)\Gamma(\delta_3)}|x_{23}^+|^{-\delta_1}|x_{13}^+|^{-\delta_2} e^{iM_1(x_{13}^2+i\epsilon)/x_{13}^+}e^{iM_2(x_{23}^2+i\epsilon)/x_{23}^+}I,\nn\\&&\label{feynmanthreeptpositionii}
\eea
where
\be
I\equiv \int_0^\infty ds_3 s_3^{\delta_3-1}|M_2+x_{12}^+s_3|^{\delta_1-1}|M_1-x_{12}^+s_3|^{\delta_2-1}e^{iv_{12}x_{12}^+s_3}\theta\left(M_1-x_{12}^+s_3\over x_{13}^+\right)\theta\left(M_2+x_{12}^+s_3\over x_{23}^+\right),\label{Iintegral}
\ee
and
\be
v_{12}\equiv {x_{12}^2+i\epsilon\over x_{12}^+}+{x_{23}^2+i\epsilon\over x_{23}^+}-{x_{13}^2+i\epsilon\over x_{13}^+}.
\ee
In order to perform the integral $I$, it helps to consider separately the eight different cases which arise when we suppose that each of $x_{12}^+$, $x_{13}^+$, and $x_{23}^+$ have definite sign. We adopt the notation $I_{ijk}$, where $i,j,k=\pm$, such that $I_{ijk}=I$ for the particular choices of sign specified by the $i,j,k$. For example $I_{++-}$ is equal to $I$ when $x_{12}^+>0$, $x_{13}^+>0$ and $x_{23}^+<0$. We need not consider $I_{+-+}$ and $I_{-+-}$ since the corresponding inequalities are not self-consistent in these cases. Furthermore, a careful inspection of (\ref{Iintegral}) reveals that it suffices to compute only the three integrals $I_{+jk}$ since the $I_{-jk}$ can be expressed in terms of these. In particular, we have
\bea
I_{-++}(\delta_1,\delta_2,\delta_3,M_1,M_2,v_{12})&=&I_{+++}(\delta_2,\delta_1,\delta_3,M_2,M_1,-v_{12}),\nn\\
I_{--+}(\delta_1,\delta_2,\delta_3,M_1,M_2,v_{12})&=&I_{++-}(\delta_2,\delta_1,\delta_3,M_2,M_1,-v_{12}),\nn\\
I_{---}(\delta_1,\delta_2,\delta_3,M_1,M_2,v_{12})&=&I_{+--}(\delta_2,\delta_1,\delta_3,M_2,M_1,-v_{12}).\label{Irelationsii}
\eea
In terms of these quantities, $I$ is given by
\bea
I&=&\theta(x_{12}^+)\theta(x_{13}^+)\theta(x_{23}^+)I_{+++}+\theta(x_{12}^+)\theta(x_{13}^+)\theta(x_{32}^+)I_{++-}+\theta(x_{12}^+)\theta(x_{31}^+)\theta(x_{32}^+)I_{+--}\nn\\&+& \theta(x_{21}^+)\theta(x_{13}^+)\theta(x_{23}^+)I_{-++}+\theta(x_{21}^+)\theta(x_{31}^+)\theta(x_{23}^+)I_{--+}+\theta(x_{21}^+)\theta(x_{31}^+)\theta(x_{32}^+)I_{---}.\nn\\&&
\eea
We begin by computing $I_{+++}$:
\be
I_{+++}=|x_{12}^+|^{-\delta_3}\int_0^\infty dz z^{\delta_3-1}(M_2+z)^{\delta_1-1}(M_1-z)^{\delta_2-1}e^{iv_{12}z}\theta(M_1-z)\theta(M_2+z).
\ee
We arrived at this expression by introducing a new integration variable via $s_3=x_{12}^+z$. To proceed further, it is useful to note that the step functions impose the constraints $M_1>0$ and $M_1+M_2>0$. We may then break up the integral as follows:
\be
I_{+++}=|x_{12}^+|^{-\delta_3}\theta(M_1)[\theta(M_2)I_{+++}^{(1)}+\theta(-M_2)\theta(M_1-|M_2|)I_{+++}^{(2)}].
\ee
We first focus on $I_{+++}^{(1)}$. When $M_2>0$, the $z$ integration limits are 0 and $M_1$. Taking $z\to M_1z$, we find
\be
I_{+++}^{(1)}=M_1^{\delta_2+\delta_3-1}M_2^{\delta_1-1}\int_0^1 dzz^{\delta_3-1}(1-z)^{\delta_2-1}\left(1+{M_1\over M_2}z\right)^{\delta_1-1}e^{iv_{12}M_1z}.
\ee
We can evaluate this integral by making use of the identity \cite{GR}
\be
\int_0^1dx x^{\nu-1}(1-x)^{\lambda-1}(1-\beta x)^{-\rho}e^{\mu x}=B(\nu,\lambda)\Phi_1(\nu,\rho,\lambda+\nu,\beta,\mu),\label{GRidentity}
\ee
where $B$ is the Euler beta function,
\be
B(x,y)=\int_0^1dt t^{x-1}(1-t)^{y-1},\label{defeulerbeta}
\ee
and $\Phi_1$ is a hypergeometric function of two variables,
\be
\Phi_1(\alpha,\beta,\gamma,x,y)=\sum_{m,n=0}^\infty{(\alpha)_{m+n}(\beta)_m\over(\gamma)_{m+n}m!n!}x^my^n.\label{defhypgeom}
\ee
This identity holds provided $\hbox{Re}(\lambda)>0$, $\hbox{Re}(\nu)>0$, and $|\hbox{arg}(1-\beta)|<\pi$. For our purposes, the first two conditions will be always be satisfied if $\delta_i>0$, while the second condition requires $\beta=-M_1/M_2<1$, which is true in the case of the integral $I_{+++}^{(1)}$, for which both $M_1$ and $M_2$ are positive. We find
\be
I_{+++}^{(1)}=M_1^{\delta_2+\delta_3-1}M_2^{\delta_2-1}B(\delta_3,\delta_2)\Phi_1(\delta_3,1-\delta_1,\delta_2+\delta_3,-{M_1\over M_2},iv_{12}M_1).
\ee
The second piece of the integral, $I_{+++}^{(2)}$, has the form
\be
I_{+++}^{(2)}=M_1^{\delta_2+\delta_3-1}|M_2|^{\delta_1-1}\int_{|M_2|\over M_1}^1dz z^{\delta_3-1}(1-z)^{\delta_2-1}\left(-1+{M_1\over|M_2|}z\right)^{\delta_1-1}e^{iv_{12}M_1z}.
\ee
We can again evaluate this by making use of the identity (\ref{GRidentity}) if we first perform the change of variable $z=(1-|M_2|/M_1)\tilde z+|M_2|/M_1$. We obtain
\bea
I_{+++}^{(2)}&=&(M_1-|M_2|)^{\delta_1+\delta_2-1}|M_2|^{\delta_3-1}e^{iv_{12}|M_2|}B(\delta_1,\delta_2)\nn\\&\times& \Phi_1(\delta_1,1-\delta_3,\delta_1+\delta_2,{|M_2|-M_1\over|M_2|},iv_{12}(M_1-|M_2|)).
\eea
Putting the results for $I_{+++}^{(1)}$ and $I_{+++}^{(2)}$ together gives
\bea
I_{+++}&=&|x_{12}^+|^{-\delta_3}\theta(M_1)\bigg\{\theta(M_2)M_1^{\delta_2+\delta_3-1}M_2^{\delta_1-1}B(\delta_3,\delta_2)\Phi_1(\delta_3, 1-\delta_1,\delta_2+\delta_3,-{M_1\over M_2},iv_{12}M_1)\nn\\&+&\theta(-M_2)\theta(M_1-|M_2|)(M_1-|M_2|)^{\delta_1+\delta_2-1}|M_2|^{\delta_3-1}e^{iv_{12}|M_2|}\nn\\&\times& B(\delta_1,\delta_2)\Phi_1(\delta_1,1-\delta_3,\delta_1+\delta_2,{|M_2|-M_1\over|M_2|},iv_{12}(M_1-|M_2|))\bigg\}.\label{Ipppii}
\eea
Similar methods lead to
\bea
I_{++-}&=&|x_{12}^+|^{-\delta_3}\int_0^\infty dzz^{\delta_3-1}|M_2+z|^{\delta_1-1}|M_1-z|^{\delta_2-1}e^{iv_{12}z}\theta(M_1-z)\theta(-M_2-z)\nn\\&=&|x_{12}^+|^{-\delta_3}\theta(M_1)\theta(-M_2)\bigg\{\theta(M_1-|M_2|)M_1^{\delta_2-1}|M_2|^{\delta_1+\delta_3-1}B(\delta_3,\delta_1)\nn\\&\times&\Phi_1(\delta_3, 1-\delta_2,\delta_1+\delta_3,{|M_2|\over M_1},iv_{12}|M_2|)+\theta(|M_2|-M_1)M_1^{\delta_2+\delta_3-1}|M_2|^{\delta_1-1}\nn\\&\times&B(\delta_3,\delta_2)\Phi_1(\delta_3,1-\delta_1,\delta_2+\delta_3,{M_1\over|M_2|},iv_{12}M_1)\bigg\}.\label{Ippmii}
\eea
\bea
I_{+--}&=&|x_{12}^+|^{-\delta_3}\int_0^\infty dzz^{\delta_3-1}|M_2-z|^{\delta_1-1}|M_1+z|^{\delta_2-1}e^{-iv_{12}z}\theta(M_1+z)\theta(M_2-z)\nn\\&=&|x_{12}^+|^{-\delta_3}\theta(-M_2)\bigg\{\theta(M_1)\theta(|M_2|-M_1)e^{iv_{12}M_1}M_1^{\delta_3-1}(|M_2|-M_1)^{\delta_1+\delta_2-1}\nn\\&\times& B(\delta_2,\delta_1)\Phi_1(\delta_2,1-\delta_3,\delta_1+\delta_2,{M_1-|M_2|\over M_1},iv_{12}(|M_2|-M_1))\nn\\&+&\theta(-M_1)|M_1|^{\delta_2-1}|M_2|^{\delta_1+\delta_3-1} B(\delta_3,\delta_1)\Phi_1(\delta_3,1-\delta_2,\delta_1+\delta_3,-{M_2\over M_1},iv_{12}|M_2|)\bigg\}.\nn\\&&\label{Ipmmii}
\eea

\section{Zero-temperature time-ordered three-point functions in momentum space}\label{appendixB}
To compute the time-ordered three-point function in momentum space, we will Fourier-transform the position space correlator. It turns out that the most efficient way to perform the Fourier transform is to start with the relativistic function in position space expressed in terms of light cone coordinates:
\bea
g^{(3)}_F&=&-\langle0|{\cal T}O_1(x_1)O_2(x_2)O_3(x_3)|0\rangle=-{1\over[x_{23}^+x_{23}^-+x_{23}^2+i\epsilon]^{\delta_1}}\nn\\&\times&{1\over[x_{13}^+x_{13}^-+x_{13}^2+i\epsilon]^{\delta_2}} {1\over[x_{12}^+x_{12}^-+x_{12}^2+i\epsilon]^{\delta_3}}.
\eea
We could start directly with the non-relativistic result (\ref{feynmanthreeptposition}), but this approach proves more cumbersome. Taking the Fourier transform with respect to all coordinates and introducing Schwinger parameters leads to
\bea
\widetilde{G}^{(3)}_F&=&-{(2\pi)^{d+2}(-i)^{\delta_1+\delta_2+\delta_3}\over8\Gamma(\delta_1)\Gamma(\delta_2)\Gamma(\delta_3)}\delta(\sum_i\Omega_i)\delta(\sum_iM_i)\delta^d(\sum_i\vec{p}_i)\int_0^\infty ds_1ds_2ds_3s_1^{\delta_1-1}s_2^{\delta_2-1}s_3^{\delta_3-1}\nn\\&\times&\int d^{d+2}x_{13}d^{d+2}x_{23}e^{i\Omega_1x_{13}^+-iM_1x_{13}^--i\vec{p}_1\cdot\vec{x}_{13}+i\Omega_2x_{23}^+-iM_2x_{23}^--i\vec{p}_2\cdot\vec{x}_{23}}\nn\\&\times&e^{is_1(x_{23}^2+x_{23}^+x_{23}^-+i\epsilon)+ is_2(x_{13}^2+x_{13}^+x_{13}^-+i\epsilon)+is_3(x_{12}^2+x_{12}^+x_{12}^-+i\epsilon)}.
\eea
The momentum variables $\Omega_i$, $M_i$ and $\vec{p}_i$ are conjugate to $x_i^+$, $x_i^-$ and $\vec{x}_i$ respectively. To obtain this expression, we have switched integration variables from $x_i$ to $x_{13}$, $x_{23}$ and $x_3$, and we subsequently computed the $d+2$ integrals over $x_3$, giving rise to the $\delta$-function factors out front. The integrals over $x_{13}^-$ and $x_{23}^-$ give rise to $\delta$-functions, and the integrals over $\vec{x}_{13}$ and $\vec{x}_{23}$ can be evaluated using
\bea
&&\int d^d x_{13}d^d x_{23} e^{i(s_2+s_3)x_{13}^2+i(s_2+s_3)x_{23}^2-2is_3\vec{x}_{13}\cdot\vec{x}_{23}-i\vec{p}_1\cdot\vec{x}_{13}-i\vec{p}_2\cdot\vec{x}_{23}}\nn\\&=& {(i\pi)^d\over(s_1s_2+s_1s_3+s_2s_3)^{d/2}}e^{-{i\over4}{s_1p_1^2+s_2p_2^2+s_3p_3^2\over s_1s_2+s_1s_3+s_2s_3}}.
\eea
We then find
\bea
\widetilde{G}^{(3)}_F&=&-{(2\pi)^{d+4}(-i)^{\delta_1+\delta_2+\delta_3}(i\pi)^d\over8\Gamma(\delta_1)\Gamma(\delta_2)\Gamma(\delta_3)}\delta(\sum_i\Omega_i)\delta(\sum_iM_i)\delta^d(\sum_i\vec{p}_i)\int_0^\infty ds_1ds_2ds_3\nn\\&\times&{s_1^{\delta_1-1}s_2^{\delta_2-1}s_3^{\delta_3-1}\over(s_1s_2+s_1s_3+s_2s_3)^{d/2}}e^{-{i\over4}{s_1p_1^2+s_2p_2^2+s_3p_3^2\over s_1s_2+s_1s_3+s_2s_3}}\int dx_{13}^+dx_{23}^+e^{i\Omega_1x_{13}^++i\Omega_2x_{23}^+}\nn\\&\times&\delta((s_2+s_3)x_{13}^+-s_3x_{23}^+-M_1)\delta(-s_3x_{13}^++(s_1+s_3)x_{23}^+-M_2).
\eea
Using the $\delta$-functions to evaluate the integrals over $x_{13}^+$ and $x_{23}^+$ yields
\bea
\widetilde{G}^{(3)}_F&=&-{(2\pi)^{d+4}(-i)^{\delta_1+\delta_2+\delta_3}(i\pi)^d\over8\Gamma(\delta_1)\Gamma(\delta_2)\Gamma(\delta_3)}\delta(\sum_i\Omega_i)\delta(\sum_iM_i)\delta^d(\sum_i\vec{p}_i)\int_0^\infty ds_1ds_2ds_3\nn\\&\times&{s_1^{\delta_1-1}s_2^{\delta_2-1}s_3^{\delta_3-1}\over(s_1s_2+s_1s_3+s_2s_3)^{{d\over2}+1}}e^{{i\over4}{(4M_1\Omega_1-p_1^2)s_1+ (4M_2\Omega_2-p_2^2)s_2+(4M_3\Omega_3-p_3^2)s_3\over s_1s_2+s_1s_3+s_2s_3}}.
\eea
To facilitate evaluation of the $s_i$ integrals, we make the following change of variables:
\be
u_i={s_1s_2+s_1s_3+s_2s_3\over s_i},
\ee
under which the integral becomes
\bea
\widetilde{G}^{(3)}_F&=&-{(2\pi)^{d+4}(-i)^{\delta_1+\delta_2+\delta_3}(i\pi)^d\over8\Gamma(\delta_1)\Gamma(\delta_2)\Gamma(\delta_3)}\delta(\sum_i\Omega_i)\delta(\sum_iM_i)\delta^d(\sum_i\vec{p}_i) \nn\\&\times&\int_0^\infty \prod_j du_j u_j^{\nu_j-1}e^{{i\over4}[4M_j\Omega_j-p_j^2+i\epsilon]/u_j} (u_1+u_2+u_3)^{-\sum_i\delta_i+{d\over2}+1}.
\eea
We have defined
\be
\nu_j\equiv \sum_{i\ne j}\delta_i - {d\over2}-1=\Delta_j-{d\over2}-1.\label{defofnuii}
\ee
Further progress is made with the help of an additional Schwinger parameter:
\bea
\widetilde{G}^{(3)}_F&=&-{(2\pi)^{d+4}(-i)^{\delta_1+\delta_2+\delta_3}(i\pi)^d\over4\Gamma(\delta_1)\Gamma(\delta_2)\Gamma(\delta_3)\Gamma(\sum_i\delta_i-{d\over2}-1)}\delta(\sum_i\Omega_i)\delta(\sum_iM_i)\delta^d(\sum_i\vec{p}_i) \nn\\&\times&\int_0^\infty dzz^{2\sum_i\delta_i-d-3}\int_0^\infty \prod_j du_j u_j^{\nu_j-1}e^{{i\over4}[4M_j\Omega_j-p_j^2+i\epsilon]/u_j} e^{-z^2u_j}.
\eea
The integral over each $u_j$ can now be identified as being essentially a modified Bessel function:
\bea
\widetilde{G}^{(3)}_F&=&{2^{5d/2+8-2\sum_i\delta_i}\pi^{2d+4}\over\Gamma(\delta_1)\Gamma(\delta_2)\Gamma(\delta_3)\Gamma(\sum_i\delta_i-{d\over2}-1)}\delta(\sum_i\Omega_i)\delta(\sum_iM_i)\delta^d(\sum_i\vec{p}_i) \nn\\&\times&\int_0^\infty dzz^{d\over2}\prod_j(-4M_j\Omega_j+p_j^2-i\epsilon)^{\nu_j/2}\mathrm{K}_{\nu_j}\left(z\sqrt{-4M_j\Omega_j+p_j^2-i\epsilon}\right).\label{feynmanthreeptmomentumii}
\eea
In this expression, we have rotated the $z$-integration contour to eliminate a factor of $\sqrt{i}$ that would otherwise appear in the argument of the Bessel function. This rotation can be done safely since the branch cut of the Bessel function lies along the negative real axis, and the square root in its argument is defined to be such that its real part is non-negative.

\section{Zero-temperature Wightman three-point functions in position space}\label{appendixC}
We begin with one of the relativistic Wightman functions in position space:
\bea
g_{123}(x_1,x_2,x_3)&\equiv&-\langle0| O_1(x_1)O_2(x_2)O_3(x_3)|0\rangle=-{1\over [x_{23}^+x_{23}^-+x_{23}^2+i\epsilon(x_{23}^+-x_{23}^-)]^{\delta_1}}\nn\\&\times&{1\over [x_{13}^+x_{13}^-+x_{13}^2+i\epsilon(x_{13}^+-x_{13}^-)]^{\delta_2}}{1\over [x_{12}^+x_{12}^-+x_{12}^2+i\epsilon(x_{12}^+-x_{12}^-)]^{\delta_3}}.\nn\\&&
\eea
We factor out $x_{ij}-i\epsilon$ from each of the three denominators and take the Fourier transform:
\bea
G_{123}&=&-(x_{23}^+-i\epsilon)^{-\delta_1}(x_{13}^+-i\epsilon)^{-\delta_2}(x_{12}^+-i\epsilon)^{-\delta_3}\int dx_1^-dx_2^-dx_3^-e^{-iM_1x_1^--iM_2x_2^--iM_3x_3^-}\nn\\&\times&\left[x_{23}^-+{x_{23}^2x_{23}^+\over(x_{23}^+)^2+\epsilon^2}+i\epsilon_1\right]^{-\delta_1} \left[x_{13}^-+{x_{13}^2x_{13}^+\over(x_{13}^+)^2+\epsilon^2}+i\epsilon_2\right]^{-\delta_2} \nn\\ &\times& \left[x_{12}^-+{x_{12}^2x_{12}^+\over(x_{12}^+)^2+\epsilon^2}+i\epsilon_3\right]^{-\delta_3}.
\eea
We have defined
\be
\epsilon_1\equiv \epsilon{(x_{23}^+)^2+x_{23}^2\over(x_{23}^+)^2+\epsilon^2},\quad \epsilon_2\equiv \epsilon{(x_{13}^+)^2+x_{13}^2\over(x_{13}^+)^2+\epsilon^2}, \quad \epsilon_3\equiv \epsilon{(x_{12}^+)^2+x_{12}^2\over(x_{12}^+)^2+\epsilon^2}.
\ee
Since the $\epsilon_i$ are positive-definite, we may introduce Schwinger parameters. After changing integration variables from $x_i^-$ to $x_{13}^-$, $x_{23}^-$ and $x_3^-$, we arrive at
\bea
G_{123}&=&-{2\pi(-i)^{\delta_1+\delta_2+\delta_3}\delta(\sum_iM_i)\over \Gamma(\delta_1)\Gamma(\delta_2)\Gamma(\delta_3)}(x_{23}^+-i\epsilon)^{-\delta_1}(x_{13}^+-i\epsilon)^{-\delta_2}(x_{12}^+-i\epsilon)^{-\delta_3}\nn\\&\times&\int_0^\infty ds_1ds_2ds_3 s_1^{\delta_1-1}s_2^{\delta_2-1}s_3^{\delta_3-1}\int dx_{13}^-dx_{23}^-e^{-iM_1x_{13}^--iM_2x_{23}^-}\nn\\&\times& e^{-s_1\left[\epsilon_1-ix_{23}^--i{x_{23}^2x_{23}^+\over(x_{23}^+)^2+\epsilon^2}\right]}e^{-s_2\left[\epsilon_2-ix_{13}^--i{x_{13}^2x_{13}^+\over(x_{13}^+)^2+\epsilon^2}\right]}
e^{-s_3\left[\epsilon_1-ix_{12}^--i{x_{12}^2x_{12}^+\over(x_{12}^+)^2+\epsilon^2}\right]}.
\eea
The integrations on $x_{13}^-$ and $x_{23}^-$ yield $\delta$-functions of the form $\delta(-M_1+s_2+s_3)$ and $\delta(-M_2+s_1-s_3)$. Since the $s_i$ only assume positive values, the first of these $\delta$-functions imposes the condition that $G_{123}$ will only be non-vanishing if $M_1>0$. This is in keeping with expectations from perturbation theory, where a negative $M_1$ would imply that $O_1(x_1)$ is comprised of creation operators. Since the creation operators will annihilate the bra vacuum state $\langle0|$, $G_{123}$ must vanish in this case. Furthermore, the two $\delta$-functions together imply that the $s_i$ integrals are supported on the curve $s_1+s_2=M_1+M_2$. The factor $\delta(M_1+M_2+M_3)$ appearing in $G_{123}$ then ensures that $M_3=-M_1-M_2<0$ when $G_{123}$ is non-vanishing. This is once again in keeping with perturbation theory if we now consider $O_3(x_3)$ acting on the ket $|0\rangle$. We also note that $M_1+M_2>0$ implies that when $M_2<0$, we must have $M_1>|M_2|$. These observations are useful for simplifying the remainder of the calculation of the Wightman function $G_{123}$, as we will see.

Evaluating the $\delta$-functions arising from the $x_{13}^-$ and $x_{23}^-$ integrations gives
\bea
G_{123}&=&-{(2\pi)^3(-i)^{\delta_1+\delta_2+\delta_3}\delta(\sum_iM_i)\over \Gamma(\delta_1)\Gamma(\delta_2)\Gamma(\delta_3)}(x_{23}^+-i\epsilon)^{-\delta_1}(x_{13}^+-i\epsilon)^{-\delta_2}(x_{12}^+-i\epsilon)^{-\delta_3}\nn\\&\times& e^{iM_1x_{13}^2{x_{13}^+\over(x_{13}^+)^2+\epsilon^2}}e^{iM_2x_{23}^2{x_{23}^+\over(x_{23}^+)^2+\epsilon^2}} \tilde I(v_{12}),
\eea
where we have defined
\be
\tilde I(v_{12})\equiv \int_0^\infty ds_3 s_3^{\delta_3-1}(M_2+s_3)^{\delta_1-1}(M_1-s_3)^{\delta_2-1}e^{i v_{12}s_3}\theta(M_1-s_3)\theta(M_2+s_3),\label{Itintegral}
\ee
and
\bea
v_{12}&\equiv& {x_{12}^2x_{12}^+\over(x_{12}^+)^2+\epsilon^2}+{x_{23}^2x_{23}^+\over(x_{23}^+)^2+\epsilon^2}-{x_{13}^2x_{13}^+\over(x_{13}^+)^2+\epsilon^2}+i(\epsilon_1-\epsilon_2+\epsilon_3) \nn\\&\approx&{x_{12}^2\over x_{12}^+-i\epsilon}+{x_{23}^2\over x_{23}^+-i\epsilon}-{x_{13}^2\over x_{13}^+-i\epsilon}.
\eea
We have already evaluated the integral $\tilde I(v_{12})$ in appendix \ref{appendixA}. It is given by
\bea
\tilde I(v_{12})&=&|x_{12}^+|^{\delta_3}I_{+++}=\theta(M_1)\bigg\{\theta(M_2)M_1^{\delta_2+\delta_3-1}M_2^{\delta_1-1}B(\delta_3,\delta_2)\nn\\&\times&\Phi_1(\delta_3, 1-\delta_1,\delta_2+\delta_3,-{M_1\over M_2},iv_{12}M_1)+\theta(-M_2)\theta(M_1-|M_2|)e^{iv_{12}|M_2|}|M_2|^{\delta_3-1}\nn\\&\times&(M_1-|M_2|)^{\delta_1+\delta_2-1} B(\delta_1,\delta_2)\Phi_1(\delta_1,1-\delta_3,\delta_1+\delta_2,{|M_2|-M_1\over|M_2|},iv_{12}(M_1-|M_2|))\bigg\}.\nn\\&&
\eea
Putting everything together, we find
\bea
G_{123}&=&-{(2\pi)^3(-i)^{\delta_1+\delta_2+\delta_3}\delta(\sum_iM_i)\over \Gamma(\delta_1)\Gamma(\delta_2)\Gamma(\delta_3)}(x_{23}^+-i\epsilon)^{-\delta_1}(x_{13}^+-i\epsilon)^{-\delta_2}(x_{12}^+-i\epsilon)^{-\delta_3}\nn\\&\times& e^{iM_1x_{13}^2{x_{13}^+\over(x_{13}^+)^2+\epsilon^2}}e^{iM_2x_{23}^2{x_{23}^+\over(x_{23}^+)^2+\epsilon^2}} \theta(M_1)\bigg\{\theta(M_2)M_1^{\delta_2+\delta_3-1}M_2^{\delta_1-1}B(\delta_3,\delta_2)\nn\\&\times&\Phi_1(\delta_3, 1-\delta_1,\delta_2+\delta_3,-{M_1\over M_2},iv_{12}M_1)+\theta(-M_2)\theta(M_1-|M_2|)e^{iv_{12}|M_2|}|M_2|^{\delta_3-1}\nn\\&\times&(M_1-|M_2|)^{\delta_1+\delta_2-1} B(\delta_1,\delta_2)\Phi_1(\delta_1,1-\delta_3,\delta_1+\delta_2,{|M_2|-M_1\over|M_2|},iv_{12}(M_1-|M_2|))\bigg\}.\nn\\&&\label{wightmanthreeptpositionii}
\eea

\section{Zero-temperature retarded and advanced three-point functions in position space}\label{appendixD}
We will compute the retarded three-point function by taking the Fourier transform of the momentum space result we obtained in section \ref{secretardedthreeptmom}. The retarded three-point function in momentum space can be written in the form
\bea
\widetilde{G}_R^{(3)}&=&{2^{5d/2+8-2\sum_i\delta_i}\pi^{2d+4}\over\Gamma(\delta_1)\Gamma(\delta_2)\Gamma(\delta_3)\Gamma(\sum_i\delta_i-{d\over2}-1)}\delta(\sum_i\Omega_i)\delta(\sum_iM_i) \delta^d(\sum_i\vec{p}_i)\int_0^\infty dz z^{d/2}\nn\\&\times&|2M_1|^{\nu_1/2}|2M_2|^{\nu_2/2}|2M_3|^{\nu_3/2}\left(-2\Omega_1+{p_1^2\over2M_1}+i\epsilon\right)^{\nu_1/2}\left(-2\Omega_2+{p_2^2\over2M_2}+i\epsilon\right)^{\nu_2/2}\nn\\&\times& \left(-2\Omega_3+{p_3^2\over2M_3}-i\epsilon\right)^{\nu_3/2}e^{-i\pi\theta(-M_1)\nu_1/2}e^{-i\pi\theta(-M_2)\nu_2/2}e^{i\pi\theta(-M_3)\nu_3/2}\nn\\&\times&\mathrm{K}_{\nu_1} \left(z\sqrt{2|M_1|}e^{-i\pi\theta(-M_1)/2}\sqrt{-2\Omega_1+{p_1^2\over2M_1}+i\epsilon}\right)\nn\\&\times&\mathrm{K}_{\nu_2} \left(z\sqrt{2|M_2|}e^{-i\pi\theta(-M_2)/2}\sqrt{-2\Omega_2+{p_2^2\over2M_2}+i\epsilon}\right)\nn\\&\times&\mathrm{K}_{\nu_3} \left(z\sqrt{2|M_3|}e^{i\pi\theta(-M_3)/2}\sqrt{-2\Omega_3+{p_3^2\over2M_3}-i\epsilon}\right).
\eea
For simplicity, we will focus on the case $M_1,M_2>0$ and $M_3<0$. Rotating the contour of the $z$-integration such that $z\to z\sqrt{-i}$, writing the Bessel functions in terms of the following integral representation:
\bea
&&|2M|^{\nu/2}\left(-2\Omega+{p^2\over2M}+i\epsilon\right)^{\nu/2}\mathrm{K}_{\nu}\left(z\sqrt{-i}\sqrt{2|M|}\sqrt{-2\Omega+{p^2\over2M}+i\epsilon}\right)\nn\\&&= 2^{\nu-1}e^{i\pi\nu/4}z^{\nu}\int_0^\infty duu^{\nu-1}e^{{i|M|\over2u}\left(-2\Omega+{p^2\over2M}+i\epsilon\right)-z^2u},
\eea
and performing the integral over $z$ leads to
\bea
\theta(M_1)\theta(M_2)\widetilde{G}_R^{(3)}&=&{2^{d+1}\pi^{2d+4}(-i)^{\sum_i\delta_i-d-2}\over\Gamma(\delta_1)\Gamma(\delta_2)\Gamma(\delta_3)}\delta(\sum_i\Omega_i)\delta(\sum_iM_i)\delta^d (\sum_i\vec{p}_i)\nn\\&\times&\int_0^\infty du_1du_2du_3 u_1^{\nu_1-1}u_2^{\nu_2-1}u_3^{\nu_3-1}(u_1+u_2+u_3)^{{d\over2}+1-\sum_i\delta_i}\nn\\&\times& e^{{iM_1\over2u_1}\left(-2\Omega_1+{p_1^2\over2M_1}+i\epsilon\right)}e^{{iM_2\over2u_2}\left(-2\Omega_2+{p_2^2\over2M_2}+i\epsilon\right)}e^{{iM_3\over2u_3} \left(-2\Omega_3+{p_3^2\over2M_3}-i\epsilon\right)}.
\eea
We want to compute the Fourier transform of this expression\footnote{We include a factor of 2 for each of the three $\Omega_i$ integrals because the non-relativistic energies are given by $2\Omega_i$.}:
\be
G_R^{(3)}={2^3\over(2\pi)^{3d+3}}\int \prod_id\Omega_i d^dp_i e^{-i\Omega_ix_i^++\vec{p}_i\cdot \vec{x}_i}\widetilde{G}_R^{(3)}.
\ee
One of the $\Omega_i$ integrals ($\Omega_3$ say) eliminates the factor of $\delta(\sum_i\Omega_i)$ appearing in $\widetilde{G}_R^{(3)}$. The remaining two $\Omega_i$ integrals then produce $\delta$-functions. Similarly, one of the $\vec{p}_i$ integrations removes the $\delta(\sum_i\vec{p}_i)$, and the other two $\vec{p}_i$ integrals are easily computed, with the result
\bea
&&\theta(M_1)\theta(M_2)G_R^{(3)}=-{(2\pi)^3i^{\sum_i\delta_i}\delta(\sum_iM_i)\over\Gamma(\delta_1)\Gamma(\delta_2)\Gamma(\delta_3)}\int_0^\infty du_1du_2du_3 e^{-{\epsilon\over2}\left({M_1\over u_1}+{M_2\over u_2}-{M_3\over u_3}\right)}\nn\\&&\times{(u_1u_2)^{\delta_3-1}(u_1u_3)^{\delta_2-1}(u_2u_3)^{\delta_1-1}\over(u_1+u_2+u_3)^{\sum_i\delta_i-1}} \delta\left(-{M_1\over u_1}+{M_3\over u_3}-x_{13}^+\right)\delta\left(-{M_2\over u_2}+{M_3\over u_3}-x_{23}^+\right)\nn\\&&\times e^{{-i\over u_1+u_2+u_3}\left[u_1u_3x_{13}^2+u_2u_3x_{23}^2+u_1u_2x_{12}^2\right]}.\label{retardedthreepointi}
\eea
Further progress is facilitated by the following change of integration variables:
\be
s_i={1\over u_i}{u_1u_2u_3\over u_1+u_2+u_3},\qquad u_i={s_1s_2+s_1s_3+s_2s_3\over s_i}.
\ee
After using the two $\delta$-functions to perform the integrations over $s_1$ and $s_2$, we are left with\footnote{The $\epsilon$ in (\ref{retardedthreepointi}) can be set to zero since the integral remains convergent in this limit. A different $\epsilon$ regulator is introduced in (\ref{retardedthreepointii}) to impose the vanishing of $G_R^{(3)}$ at $x_{13}^+=0$ and $x_{23}^+=0$. The fact that $G_R^{(3)}$ vanishes at these points is evident from (\ref{retardedthreepointi}) since one or both of the $\delta$-functions vanish identically.}:
\bea
&&\theta(M_1)\theta(M_2)G_R^{(3)}=-{(2\pi)^3i^{\sum_i\delta_i}\delta(\sum_iM_i)\over\Gamma(\delta_1)\Gamma(\delta_2)\Gamma(\delta_3)}|x_{13}^+|^{-\delta_2}|x_{23}^+|^{-\delta_1}e^{iM_1 (x_{13}^2+i\epsilon)/x_{13}^+}\nn\\&&\times e^{iM_2(x_{23}^2+i\epsilon)/x_{23}^+}\int_0^\infty ds_3s_3^{\delta_3-1}|-M_2+x_{12}^+s_3|^{\delta_1-1}|-M_1-x_{12}^+s_3|^{\delta_2-1} e^{-iv_{12}x_{12}^+s_3}\nn\\&&\times\theta\left({-M_2+x_{12}^+s_3\over x_{23}^+}\right)\theta\left(-M_1-x_{12}^+s_3\over x_{13}^+\right),\label{retardedthreepointii}
\eea
where
\be
v_{12}= {x_{12}^2\over x_{12}^+}+{x_{23}^2\over x_{23}^+}-{x_{13}^2\over x_{13}^+}.
\ee
Comparing the integral in (\ref{retardedthreepointii}) with (\ref{Iintegral}), we see that we have essentially already evaluated this integral. It can be expressed in terms of the functions $I_{+++}$, $I_{++-}$, and $I_{+--}$ we defined in section \ref{secfeynmanthreeptpos} and appendix \ref{appendixA}. We must first send $M_1\to-M_1$, $M_2\to-M_2$, and $v_{12}\to-v_{12}$ and then impose the constraints $M_1>0$, $M_2>0$. From (\ref{Ipppii}), (\ref{Ippmii}), and (\ref{Ipmmii}), we see that only $I_{+--}$ is non-vanishing under these constraints. Since $I_{---}$ can be obtained from $I_{+--}$ by swapping parameters (see (\ref{Irelationsii})), this function will also contribute, and we find
\bea
\theta(M_1)\theta(M_2)G_R^{(3)}&=&-{(2\pi)^3i^{\sum_i\delta_i}\delta(\sum_iM_i)\over\Gamma(\delta_1)\Gamma(\delta_2)\Gamma(\delta_3)}|x_{13}^+|^{-\delta_2} |x_{23}^+|^{-\delta_1} |x_{12}^+|^{-\delta_3} e^{iM_1 (x_{13}^2+i\epsilon)/x_{13}^+}\nn\\&\times&e^{iM_2(x_{23}^2+i\epsilon)/x_{23}^+}\theta(M_1)\theta(M_2)\theta(x_{31}^+)\theta(x_{32}^+)\bigg\{\theta(x_{12}^+) M_1^{\delta_2-1}M_2^{\delta_1+\delta_3-1}\nn\\&\times& B(\delta_3,\delta_1)\Phi_1(\delta_3,1-\delta_2,\delta_1+\delta_3,-{M_2\over M_1},-iv_{12}M_2)\nn\\&+&\theta(x_{21}^+)M_1^{\delta_2+\delta_3-1}M_2^{\delta_1-1}B(\delta_3,\delta_2)\Phi_1(\delta_3,1-\delta_1,\delta_2+\delta_3,-{M_1\over M_2},iv_{12}M_1)\bigg\}.\nn\\&&\label{retardedthreeptpositionii}
\eea
Notice that this expression is non-vanishing only when $x_3^+>x_2^+$ and $x_3^+>x_1^+$, which is consistent with our expectation that $G_R^{(3)}$ is a retarded correlator for which $x_3^+/2$ is the largest time.

The advanced three-point function can be obtained by starting from $\widetilde{G}_A^{(3)}=(\widetilde{G}_R^{(3)})^*$. We again suppose that $M_1,M_2>0$, $M_3<0$. Similar steps to the above lead to the expression
\bea
&&\theta(M_1)\theta(M_2)G_A^{(3)}=-{(2\pi)^3(-i)^{\sum_i\delta_i}\delta(\sum_iM_i)\over\Gamma(\delta_1)\Gamma(\delta_2)\Gamma(\delta_3)}|x_{13}^+|^{-\delta_2}|x_{23}^+|^{-\delta_1}e^{iM_1 (x_{13}^2+i\epsilon)/x_{13}^+}\nn\\&&\times e^{iM_2(x_{23}^2+i\epsilon)/x_{23}^+}\int_0^\infty ds_3s_3^{\delta_3-1}|M_2+x_{12}^+s_3|^{\delta_1-1}|M_1-x_{12}^+s_3|^{\delta_2-1} e^{iv_{12}x_{12}^+s_3}\nn\\&&\times\theta\left({M_2+x_{12}^+s_3\over x_{23}^+}\right)\theta\left(M_1-x_{12}^+s_3\over x_{13}^+\right).\label{advancedthreepointii}
\eea
This time, the integral over $s_3$ is precisely the same as $I$ as it is given in (\ref{Iintegral}). The conditions $M_1>0$ and $M_2>0$ now single out $I_{+++}$ and $I_{-++}$ as the sole contributors, and we find
\bea
\theta(M_1)\theta(M_2)G_A^{(3)}&=&-{(2\pi)^3(-i)^{\sum_i\delta_i}\delta(\sum_iM_i)\over\Gamma(\delta_1)\Gamma(\delta_2)\Gamma(\delta_3)}|x_{13}^+|^{-\delta_2} |x_{23}^+|^{-\delta_1} |x_{12}^+|^{-\delta_3} e^{iM_1 (x_{13}^2+i\epsilon)/x_{13}^+}\nn\\&\times&e^{iM_2(x_{23}^2+i\epsilon)/x_{23}^+}\theta(M_1)\theta(M_2)\theta(x_{13}^+)\theta(x_{23}^+)\bigg\{\theta(x_{12}^+) M_1^{\delta_2+\delta_3-1}M_2^{\delta_1-1}\nn\\&\times& B(\delta_3,\delta_2)\Phi_1(\delta_3,1-\delta_1,\delta_2+\delta_3,-{M_1\over M_2},iv_{12}M_1)\nn\\&+&\theta(x_{21}^+)M_1^{\delta_2-1}M_2^{\delta_1+\delta_3-1}B(\delta_3,\delta_1)\Phi_1(\delta_3,1-\delta_2,\delta_1+\delta_3,-{M_2\over M_1},-iv_{12}M_2)\bigg\}.\nn\\&&\label{advancedthreeptpositionii}
\eea
We see that this result is consistent with the interpretation as an advanced correlator, with $x_3^+/2$ now being the smallest time.

\section{$\delta {g_1}^2$ metric fluctuation spectrum}\label{appendixE}

Let us begin with a brief reminder of the scalar spherical harmonics on $S^5$, written as an $S^1$ fibration over $CP^2$:
\bea
ds_{S^5}^2&=&(d\chi+\sin^2\mu \,\omega_3)^2+ [d\mu^2+\sin^2\mu(\omega_1^2+\omega_2^2+\cos^2\mu\,\omega_3^2)]\nn\\
\omega_1&=&\frac12(\cos\psi\, d\theta+\sin\psi\sin\theta\,d\phi)\nn\\
\omega_2&=&=\frac 12(\sin\psi\,d\theta-\cos\psi \sin\theta\,d\phi)\nn\\
\omega_3&=&\frac 12(d\psi+\cos\theta\,d\phi)
\eea
where $0\leq\chi\leq 2\pi$, $0\leq \mu\leq \frac{\pi}2$, $0\leq\psi\leq 4\pi$,
$0\leq \theta\leq \pi$ and $0\leq\phi\leq 2\pi$.

The scalar spherical harmonics are solutions to the box equation
\be
\Box_{S^5} Y=\frac{1}{\sqrt{g}}\partial_i\sqrt{g} g^{ij}\partial_j Y=-l(l+4) Y,\qquad l=0,1,2\dots
\ee
where
\be
Y=e^{in_{\chi} \chi}e^{i n_{\phi} \phi}e^{i n_{\psi}\psi} \Theta(\theta) U(\mu)
\ee
and
\bea
\Theta(\theta)&=&(1-\cos\theta)^{\frac 12|n_{\psi}-n_{\phi}|}(1+\cos\theta)^{\frac 12|n_{\psi}+n_{\phi}|}{}_2F_1\bigg(-n_{\theta}, n_{\theta}+2|n_{\psi}|+1;1+|n_{\psi}+n_{\phi}|;\frac{1+\cos\theta}{2}\bigg),\nn\\
U(\mu)&=&(\sin\mu)^{2|n_{\psi}|+2n_{\theta}}(\cos\mu)^{|n_{\chi}-2n_{\psi}|}
{}_2F_1(-n_{\mu}, l-n_{\mu}+2; 1+|n_{\chi}-2n_{\psi}|;\cos^2\mu)\nn\\
l&=&|n_{\chi}-2n_{\psi}|+2|n_{\psi}|+2n_{\theta}+2n_{\mu}
\eea
and where $n_{\phi}\leq |n_{\psi}|$.

In the black hole background \cite{HRR,ABM,MMT}
\bea
ds_{10d,E}^2&=&k(r)^{\frac 14}\bigg[r^2\bigg(-\frac{\beta^2 r^2 f(r)}{k(r)}(dt+dy)^2
-\frac{f(r)}{k(r)}dt^2+\frac{dy^2}{k(r)}+d\vec x^2\bigg)
+\frac{dr^2}{r^2 f(r)}\nn\\&+&
\frac{(d\chi+\sin^2\mu \,\omega_3)^2}{k(r)}+ds_{CP^2}^2\bigg]
,\eea
one of the simplest fluctuations, which decouples from the rest, is  the metric fluctuation $\delta {g_1}^2\equiv g^{22}\delta g_{12}(t,y,\vec x,r) Y(\chi,\mu,\psi,\theta,\phi)$, which in the Einstein frame obeys the equation of motion
\be
\Box_{10d,E} {\delta g_1}^2=0\label{10deqn}.
\ee
The other Einstein equations impose additional constraints, namely ${\delta g_1}^2$ must be independent of $x_1, x_2$ coordinates.
Substituting the spherical harmonics and the black hole metric into (\ref{10deqn}) leads to
\be
\bigg[\Box_{5d,KK} - k(r)^{-\frac 13}\bigg(l(l+4)+(k(r)-1)n_{\chi}^2\bigg)\bigg]{\delta g_1}^2=0,
\label{massiveEOM}
\ee
where the D'Alembert operator $\Box_{5d,KK}$ is written with respect to the appropriately rescaled non-compact space metric, such that the 5d Einstein-Hilbert action is properly normalized upon performing a standard Kaluza-Klein reduction on the black hole background:
\be
ds_{5d,KK}^2=k(r)^{\frac 13}\bigg[
r^2\bigg(-\frac{\beta^2 r^2 f(r)}{k(r)}(dt+dy)^2
-\frac{f(r)}{k(r)}dt^2+\frac{dy^2}{k(r)}+d\vec x^2\bigg)
+\frac{dr^2}{r^2 f(r)}\bigg].
 \ee
In momentum space, and in Minkwoski signature the equation of motion for the metric fluctuation reads
\bea
{\delta g_1}^2&\equiv& F(\omega, p_y,\vec p=0;u),\nn\\
F&\equiv& u^\lambda (1-u)^{\frac{-i\omega}{2}} (1+u)^{\frac{\omega}{2}} H(u),
\nn\\
&&H''+\bigg(\frac{\gamma}{u}+\frac{\delta}{u-1}+\frac{\varepsilon}{u+1}\bigg)H'+
\frac{\alpha\beta u -q}{u(u-1)(u+1)}H=0,\nn\\
&&\lambda=1-\sqrt{1+\Lambda^2(\p_y+\gomega)^2+\frac{l(l+4)}4},\nn\\
&&\Lambda\equiv \beta r_+,\qquad \gomega\equiv\frac{\omega}{2 r_+}, \qquad \p_y\equiv\frac{p_y}{2 r_+},\label{massiveHeun}
\eea
where the $H$-equation is of Heun type, with parameters
\bea
&&\gamma=1-2\sqrt{1+\Lambda^2(\p_y+\gomega)^2+\frac{l(l+4)}4},\quad \delta=1-i\gomega_E,\quad \varepsilon=1+\gomega,\nn\\ &&\alpha=\beta={1-i\over2}\gomega+1-\sqrt{1+\Lambda^2(\p_y+\gomega)^2+\frac{l(l+4)}4},\nn\\ && q=-{1+i\over2}\gomega\left[2\sqrt{1+\Lambda^2(\p_y+\gomega)^2 +\frac{l(l+4)}4}-1\right]+\gomega^2-\p_y^2-\frac{n_{\chi}^2 \Lambda^2}4.
\eea
(We hope that the parameter of the Heun function, $\beta$, is not confused with the parameter $\beta$ of the black hole metric, which is now appearing through $\Lambda$.)

In (\ref{massiveHeun}) we have retained only the solution which behaves like an incoming wave at the horizon. The case of the trivial spherical harmonic ($l=0$) was discussed in \cite{HRR,ABM}, and was the template for the minimally coupled massless scalar fluctuation equation of motion (of course, the metric fluctuations transform as a rank 2 symmetric tensor, and this is the reason for the additional constraints which were placed on ${\delta g_1}^2$). Here we are interested in the equation of motion obeyed by massive fluctuations in the black hole background, which is why we considered the most general type of scalar spherical harmonic, leading to the equation of motion (\ref{massiveEOM}). Since the $S^5$ factor is squashed in the 10d black hole geometry, both $l$ and $n_{\chi}$ quantum numbers appear in (\ref{massiveEOM}). However, it is reassuring that  despite the complexity of the metric and of the peculiarities of the equation of motion for the massive fluctuations, the solution can still be obtained through Heun's functions.

\end{appendix}


\begin{thebibliography}{99}
\bibitem{Maldacena:1997re}
  J.~M.~Maldacena,
  ``The large N limit of superconformal field theories and supergravity,''
  Adv.\ Theor.\ Math.\ Phys.\  {\bf 2}, 231 (1998)
  [Int.\ J.\ Theor.\ Phys.\  {\bf 38}, 1113 (1999)]
  [arXiv:hep-th/9711200].
\bibitem{Witten:1998qj}
  E.~Witten,
  ``Anti-de Sitter space and holography,''
  Adv.\ Theor.\ Math.\ Phys.\  {\bf 2}, 253 (1998)
  [arXiv:hep-th/9802150].
\bibitem{Gubser:1998bc}
  S.~S.~Gubser, I.~R.~Klebanov and A.~M.~Polyakov,
  ``Gauge theory correlators from non-critical string theory,''
  Phys.\ Lett.\  B {\bf 428}, 105 (1998)
  [arXiv:hep-th/9802109].

\bibitem{OHara}
  K.~M.~O'Hara {\sl et al.},
  ``Observation of a strongly-interacting degenerate Fermi gas of atoms,"
  Science {\bf 298}, 2179 (2002).

\bibitem{Nishida:2007pj}
  Y.~Nishida and D.~T.~Son,
  ``Nonrelativistic conformal field theories,''
  Phys.\ Rev.\  D {\bf 76}, 086004 (2007)
  [arXiv:0706.3746 [hep-th]].
\bibitem{Nishida:2007mr}
  Y.~Nishida, D.~T.~Son and S.~Tan,
  ``Universal Fermi Gas with Two- and Three-Body Resonances,''
  Phys.\ Rev.\ Lett.\  {\bf 100}, 090405 (2008)
  [arXiv:0711.1562 [cond-mat.other]].
\bibitem{Son:2010kq}
  D.~T.~Son and E.~G.~Thompson,
  ``Short-distance and short-time structure of a unitary Fermi gas,''
  arXiv:1002.0922 [cond-mat.quant-gas].
\bibitem{Nishida:2010tm}
  Y.~Nishida and D.~T.~Son,
  ``Unitary Fermi gas, epsilon expansion, and nonrelativistic conformal field
  theories,''
  arXiv:1004.3597 [cond-mat.quant-gas].

\bibitem{Son}
  D.~T.~Son,
  ``Toward an AdS/cold atoms correspondence: a geometric realization of the
  Schroedinger symmetry,''
  Phys.\ Rev.\  D {\bf 78}, 046003 (2008)
  [arXiv:0804.3972 [hep-th]].
\bibitem{BM}
  K.~Balasubramanian and J.~McGreevy,
  ``Gravity duals for non-relativistic CFTs,''
  Phys.\ Rev.\ Lett.\  {\bf 101}, 061601 (2008)
  [arXiv:0804.4053 [hep-th]].

\bibitem{HRR}
  C.~P.~Herzog, M.~Rangamani and S.~F.~Ross,
  ``Heating up Galilean holography,''
  JHEP {\bf 0811}, 080 (2008)
  [arXiv:0807.1099 [hep-th]].
\bibitem{ABM}
  A.~Adams, K.~Balasubramanian and J.~McGreevy,
  ``Hot Spacetimes for Cold Atoms,''
  JHEP {\bf 0811}, 059 (2008)
  [arXiv:0807.1111 [hep-th]].
\bibitem{MMT}
  J.~Maldacena, D.~Martelli and Y.~Tachikawa,
  ``Comments on string theory backgrounds with non-relativistic conformal
  symmetry,''
  JHEP {\bf 0810}, 072 (2008)
  [arXiv:0807.1100 [hep-th]].

\bibitem{Duvali}
  C.~Duval, G.~Burdet, H.~P.~Kunzle and M.~Perrin,
  ``Bargmann Structures And Newton-Cartan Theory,''
  Phys.\ Rev.\  D {\bf 31}, 1841 (1985).
\bibitem{Duvalii}
  C.~Duval, G.~W.~Gibbons and P.~Horvathy,
  ``Celestial Mechanics, Conformal Structures, and Gravitational Waves,''
  Phys.\ Rev.\  D {\bf 43}, 3907 (1991)
  [arXiv:hep-th/0512188].
\bibitem{Duvaliii}
  C.~Duval, M.~Hassaine and P.~A.~Horvathy,
  ``The geometry of Schr\'odinger symmetry in gravity
  background/non-relativistic CFT,''
  Annals Phys.\  {\bf 324}, 1158 (2009)
  [arXiv:0809.3128 [hep-th]].

\bibitem{Alishahiha:2003ru}
  M.~Alishahiha and O.~J.~Ganor,
  ``Twisted backgrounds, pp-waves and nonlocal field theories,''
  JHEP {\bf 0303}, 006 (2003)
  [arXiv:hep-th/0301080].
\bibitem{Bergman:2000cw}
  A.~Bergman and O.~J.~Ganor,
  ``Dipoles, twists and noncommutative gauge theory,''
  JHEP {\bf 0010}, 018 (2000)
  [arXiv:hep-th/0008030].
\bibitem{Dasgupta:2000ry}
  K.~Dasgupta, O.~J.~Ganor and G.~Rajesh,
  ``Vector deformations of N = 4 super-Yang-Mills theory, pinned branes,  and
  arched strings,''
  JHEP {\bf 0104}, 034 (2001)
  [arXiv:hep-th/0010072].

\bibitem{Henkel}
  M.~Henkel,
  ``Schrodinger invariance in strongly anisotropic critical systems,''
  J.\ Statist.\ Phys.\  {\bf 75}, 1023 (1994)
  [arXiv:hep-th/9310081].
\bibitem{HU}
  M.~Henkel and J.~Unterberger,
  ``Schroedinger invariance and space-time symmetries,''
  Nucl.\ Phys.\  B {\bf 660}, 407 (2003)
  [arXiv:hep-th/0302187].

\bibitem{FM}
  C.~A.~Fuertes and S.~Moroz,
  ``Correlation functions in the non-relativistic AdS/CFT correspondence,''
  Phys.\ Rev.\  D {\bf 79}, 106004 (2009)
  [arXiv:0903.1844 [hep-th]].
\bibitem{VW}
  A.~Volovich and C.~Wen,
  ``Correlation Functions in Non-Relativistic Holography,''
  JHEP {\bf 0905}, 087 (2009)
  [arXiv:0903.2455 [hep-th]].
\bibitem{LH}
  R.~G.~Leigh and N.~N.~Hoang,
  ``Real-Time Correlators and Non-Relativistic Holography,''
  JHEP {\bf 0911}, 010 (2009)
  [arXiv:0904.4270 [hep-th]].

\bibitem{SonStarinets}
  D.~T.~Son and A.~O.~Starinets,
  ``Minkowski-space correlators in AdS/CFT correspondence: Recipe and
  applications,''
  JHEP {\bf 0209}, 042 (2002)
  [arXiv:hep-th/0205051].
\bibitem{HerzogSon}
  C.~P.~Herzog and D.~T.~Son,
  ``Schwinger-Keldysh propagators from AdS/CFT correspondence,''
  JHEP {\bf 0303}, 046 (2003)
  [arXiv:hep-th/0212072].

\bibitem{Skenderisi}
  K.~Skenderis and B.~C.~van Rees,
  ``Real-time gauge/gravity duality,''
  Phys.\ Rev.\ Lett.\  {\bf 101}, 081601 (2008)
  [arXiv:0805.0150 [hep-th]].
\bibitem{Skenderisii}
  K.~Skenderis and B.~C.~van Rees,
  ``Real-time gauge/gravity duality: Prescription, Renormalization and
  Examples,''
  JHEP {\bf 0905}, 085 (2009)
  [arXiv:0812.2909 [hep-th]].

\bibitem{BVWA}
  E.~Barnes, D.~Vaman, C.~Wu and P.~Arnold,
  ``Real-time finite-temperature correlators from AdS/CFT,''
  arXiv:1004.1179 [hep-th].

\bibitem{Blau:2010fh}
  M.~Blau, J.~Hartong and B.~Rollier,
  ``Geometry of Schroedinger Space-Times II: Particle and Field Probes of the
  Causal Structure,''
  JHEP {\bf 1007}, 069 (2010)
  [arXiv:1005.0760 [hep-th]].

\bibitem{Veltman}
  M.~J.~G.~Veltman,
  ``Unitarity and causality in a renormalizable field theory with unstable
  particles,''
  Physica {\bf 29}, 186 (1963).
\bibitem{'tHooft}
  G.~'t Hooft and M.~J.~G.~Veltman,
  ``Diagrammar,''
  NATO Adv.\ Study Inst.\ Ser.\ B Phys.\  {\bf 4}, 177 (1974).

\bibitem{kobes}
R.~Kobes, ``A Correspondence between imaginary-time and real-time finite temperature field theory'', Phys.~Rev.~D {\bf 42}, 562 (1990)

\bibitem{wightman}
A.~S.~Wightman, ``Quantum Field Theory in Terms of Vacuum Expectation Values'', Phys.~Rev. {\bf 101}, 860 (1956)
\bibitem{Hofman}
  D.~M.~Hofman and J.~Maldacena,
  ``Conformal collider physics: Energy and charge correlations,''
  JHEP {\bf 0805}, 012 (2008)
  [arXiv:0803.1467 [hep-th]].

\bibitem{ls}
H.~Lehmann, K.~Symanzik, ``On the formulation of Quantized Field Theories'', Nuovo Cim. {\bf 6}, 319 (1957)

\bibitem{192}
R.~S.~Maier, ``The 192 Solutions of the Heun Equation,''
Math. Comp. {\bf 76} 811 (2007)
[arXiv:math/0408317].

\bibitem{Liu:2010sa}
  J.~T.~Liu, P.~Szepietowski and Z.~Zhao,
  ``Consistent massive truncations of IIB supergravity on Sasaki-Einstein
  manifolds,''
  arXiv:1003.5374 [hep-th].
\bibitem{Gauntlett:2010vu}
  J.~P.~Gauntlett and O.~Varela,
  JHEP {\bf 1006}, 081 (2010)
  [arXiv:1003.5642 [hep-th]].
\bibitem{Skenderis:2010vz}
  K.~Skenderis, M.~Taylor and D.~Tsimpis,
  JHEP {\bf 1006}, 025 (2010)
  [arXiv:1003.5657 [hep-th]].
\bibitem{Cassani:2010uw}
  D.~Cassani, G.~Dall'Agata and A.~F.~Faedo,
  JHEP {\bf 1005}, 094 (2010)
  [arXiv:1003.4283 [hep-th]].
\bibitem{vanRees:2009rw}
  B.~C.~van Rees,
  ``Real-time gauge/gravity duality and ingoing boundary conditions,''
  Nucl.\ Phys.\ Proc.\ Suppl.\  {\bf 192-193} (2009) 193
  [arXiv:0902.4010 [hep-th]].


\bibitem{GR}
I.~S.~Gradshteyn and I.~M.~Ryzhik, ``Table of integrals, series, and products", fifth edition, A. Jeffrey ed.,
Academic Press, U.S.A. (1994), pg. 367

\bibitem{Leiva:2003kd}
  C.~Leiva and M.~S.~Plyushchay,
  ``Conformal symmetry of relativistic and nonrelativistic systems and  AdS/CFT
  correspondence,''
  Annals Phys.\  {\bf 307}, 372 (2003)
  [arXiv:hep-th/0301244].



\end{thebibliography}
\end{document}